\pdfoutput=1

\documentclass[11pt]{article}

\usepackage[final]{acl}

\usepackage{times}
\usepackage{latexsym}

\usepackage[T1]{fontenc}

\usepackage[utf8]{inputenc}

\usepackage{microtype}

\usepackage{inconsolata}

\usepackage{algorithm}
\usepackage{algorithmic}
\usepackage{multirow}
\usepackage{booktabs}
\usepackage{amsmath}
\usepackage{amssymb}
\usepackage{graphicx}

\usepackage{newfloat}
\usepackage{listings}

\title{Versatile Framework for Song Generation with Prompt-based Control}

\author{%
\normalsize
Yu Zhang\thanks{Equal contribution}\quad
Wenxiang Guo\footnotemark[1]\quad
Changhao Pan\footnotemark[1]\quad
Zhiyuan Zhu\footnotemark[1]\quad
Ruiqi Li\quad
Jingyu Lu\quad
\\\normalsize
\textbf{Rongjie Huang}\quad
\textbf{Ruiyuan Zhang}\quad
\textbf{Zhiqing Hong}\quad
\textbf{Ziyue Jiang}\quad
\textbf{Zhou Zhao}\thanks{Corresponding Author}\quad
\\
Zhejiang University\\
\texttt{\{yuzhang34,zhaozhou\}@zju.edu.cn}
}

\begin{document}
\maketitle
\begin{abstract}

Song generation focuses on producing controllable high-quality songs based on various prompts. 
However, existing methods struggle to generate vocals and accompaniments with prompt-based control and proper alignment. 
Additionally, they fall short in supporting various tasks.
To address these challenges, we introduce VersBand, a multi-task song generation framework for synthesizing high-quality, aligned songs with prompt-based control. 
VersBand comprises these primary models:
1) VocalBand, a decoupled model, leverages the flow-matching method for generating singing styles, pitches, and mel-spectrograms, allowing fast, high-quality vocal generation with style control.
2) AccompBand, a flow-based transformer model, incorporates the Band-MOE, selecting suitable experts for enhanced quality, alignment, and control.
This model allows for generating controllable, high-quality accompaniments aligned with vocals.
3) Two generation models, LyricBand for lyrics and MelodyBand for melodies, contribute to the comprehensive multi-task song generation system, allowing for extensive control based on multiple prompts.
Experimental results show that VersBand outperforms baseline models across multiple song generation tasks using objective and subjective metrics.
Demos and codes are available at \url{https://aaronz345.github.io/VersBandDemo} and \url{https://github.com/AaronZ345/VersBand}.

\end{abstract}
\section{Introduction}

Song generation focuses on producing complete musical pieces based on text prompts (about lyrics, melodies, singing, and music styles) and optional audio prompts. 
Unlike singing voice synthesis (SVS) \citep{zhang2024tcsinger}, which focuses on the vocal component, or music generation \citep{dong2018musegan} for synthesizing only instrumental tracks, song generation involves synthesizing both high-quality vocals and accompaniments with high-level prompt-based control and proper alignment in rhythm, melody, and beats \cite{li2024accompanied}.

\begin{figure*}[ht]
\centering
\includegraphics[width=0.95\textwidth]{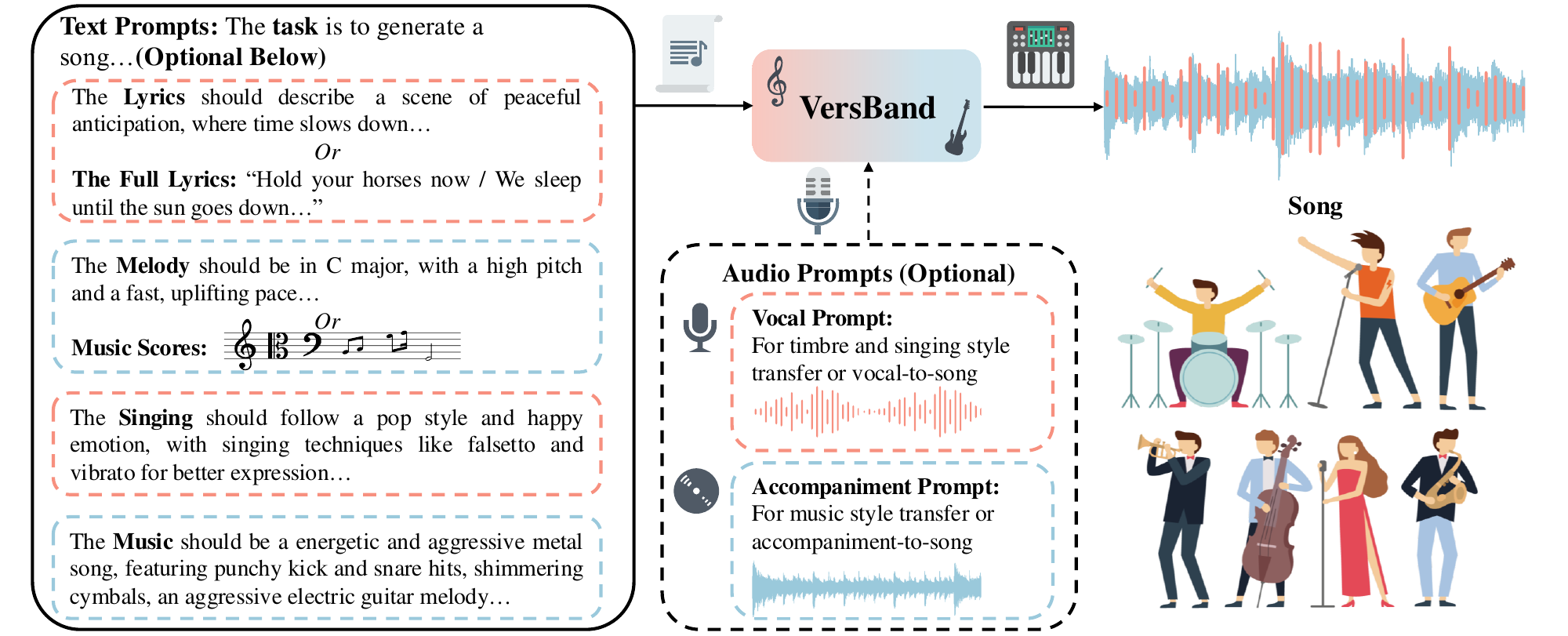}
\caption{Overview of VersBand, which generates complete songs like a versatile band.
The dashed lines indicate optional inputs.
At a minimum, users can just input "The task is to generate a song."}
\label{fig: intro}
\end{figure*}

Despite significant advancements in SVS and music domains, generating high-quality, controllable, aligned songs remains challenging. 
Song generation aims to enable controllable musical experiences, with broad applications ranging from entertainment videos to professional composition. 
As shown in Figure \ref{fig: intro}, song generation models can leverage different prompts for multiple tasks.  
\textbf{Text prompts} allow for control over tasks, lyrics, melody, singing styles (like singing methods, emotion, and techniques), and music styles (like genre, tone, and instrumentation), while \textbf{audio prompts} enable users to input their voice or preferred music for customization.
However, the few existing song generation models \citep{zhiqing2024text} lack mechanisms to properly align vocals with accompaniments and fail to achieve effective control.

Currently, song generation encounters three major challenges:
1) \textbf{Limitations in high-quality vocal generation with style control.} 
For singing style control, StyleSinger \citep{zhang2024stylesinger} conducts style transfer, while PromptSinger \citep{wang2024prompt} achieves singer identity control. 
However, existing models have yet to generate pleasing vocals with high-level style control (like singing methods, emotion, and techniques) by text prompts, and customization with audio prompts.
2) \textbf{Difficulties in controllable and aligned accompaniment generation.} 
Existing music generation models \citep{dong2018musegan} and text-to-song models \citep{zhiqing2024text} lack mechanisms for style control and aligning vocals with accompaniments in rhythm, melody, and beats. 
Generating controllable (like genre, tone, and instrumentation) and aligned accompaniments remains challenging.
3) \textbf{Challenges in multi-task song generation based on various prompts.} 
The limited existing song generation methods \citep{li2024accompanied} do not support diverse related tasks. 
This reliance on constrained inputs leads to a suboptimal user experience and restricts the models' ability to customize songs.

To address these challenges, we introduce VersBand, a multi-task song generation framework for synthesizing high-quality, aligned songs with prompt-based control.
Following the perception that accompaniment complements vocal melody with complex harmonic and rhythmic structure \citep{zhiqing2024text}, we generate them separately.
To achieve fast and high-quality vocal generation with control, we design a decoupled model, VocalBand, predicting singing styles, pitches, and mel-spectrograms based on the flow-matching method.
Based on the complex nature of music, we introduce a flow-based transformer model, AccompBand, to generate high-fidelity, controllable, aligned accompaniments.
We design Band-MOE (Mixture of Experts), selecting suitable experts for enhanced quality, alignment, and control. 
Additionally, we add two generation models, LyricBand for lyric generation and MelodyBand for melody generation, contributing to the comprehensive multi-task song generation system. 
Our experiments show that VersBand can generate high-quality songs with control, outperforming other baseline models in multiple related song generation tasks.
The main contributions are summarized as follows:

\begin{itemize}
\item We propose VersBand, a multi-task song generation approach for generating high-quality, aligned songs with prompt-based control.
\item We design a decoupled model VocalBand, which leverages the flow-matching method to generate singing styles, pitches, and mel-spectrograms, enabling fast and high-quality vocal synthesis with high-level style control.
\item We introduce a flow-based transformer model AccompBand to generate high-quality, controllable, aligned accompaniments, with the Band-MOE, selecting suitable experts for enhanced quality, alignment, and control.
\item Experimental results demonstrate that VersBand achieves superior objective and subjective evaluations compared to baseline models across multiple song generation tasks.
\end{itemize}
\section{Background}

\subsection{Singing Voice Synthesis}

Singing Voice Synthesis (SVS) rapidly advances for generating singing voices from given lyrics and music scores. 
\citet{choi2022melody} presents a melody-unsupervised model, eliminating the need for temporal alignment. 
VISinger 2 \citep{zhang2022visinger} employs digital signal processing techniques to enhance fidelity, while \citet{kim2024adversarial} uses adversarial multi-task learning to improve the singing naturalness.
StyleSinger \citep{zhang2024stylesinger} facilitates style transfer by extracting styles via a residual quantization method. 
GTSinger \citep{zhang2024gtsinger} releases a dataset with multiple style annotations.
Despite these advancements, these approaches can not generate aligned accompaniment. 
Melodist \citep{zhiqing2024text} has introduced a text-to-song model that sequentially generates vocals and accompaniments using auto-regressive transformers. 
TCSinger \citep{zhang2024tcsinger} and TechSinger \citep{guo2025techsinger} utilize input sequences to control vocal techniques.
However, achieving high-quality vocal generation with high-level control by natural language text or audio prompts remains a challenging task.

\subsection{Accompaniment Generation}

Research in accompaniment usually focuses on musical symbolic tokens. 
MuseGAN \citep{dong2018musegan} employs a GAN-based approach to generate symbolic music. 
SongMASS \citep{sheng2021songmass} uses transformer models to generate lyrics or melodies conditioned on each other. 
MusicLM \citep{agostinelli2023musiclm} leverages joint textual-music representations from MuLan \citep{huang2022mulan} to generate semantic and acoustic tokens using transformer decoders. 
MusicLDM \citep{chen2024musicldm} incorporates beat-tracking information to address potential plagiarism concerns in music generation. 
SongCreator \citep{lei2024songcreator} can create songs based on lyrics and audio prompts.
Recently, MelodyLM \citep{li2024accompanied} has employed transformer and diffusion models for decoupled song generation. 
Nevertheless, challenges remain in generating high-quality music with effective control based on multiple prompts. 
Existing methods also lack mechanisms for alignment with vocals and support for multiple song generation tasks.

\section{Method}

\begin{figure*}[ht]
\centering
\includegraphics[width=1\textwidth]{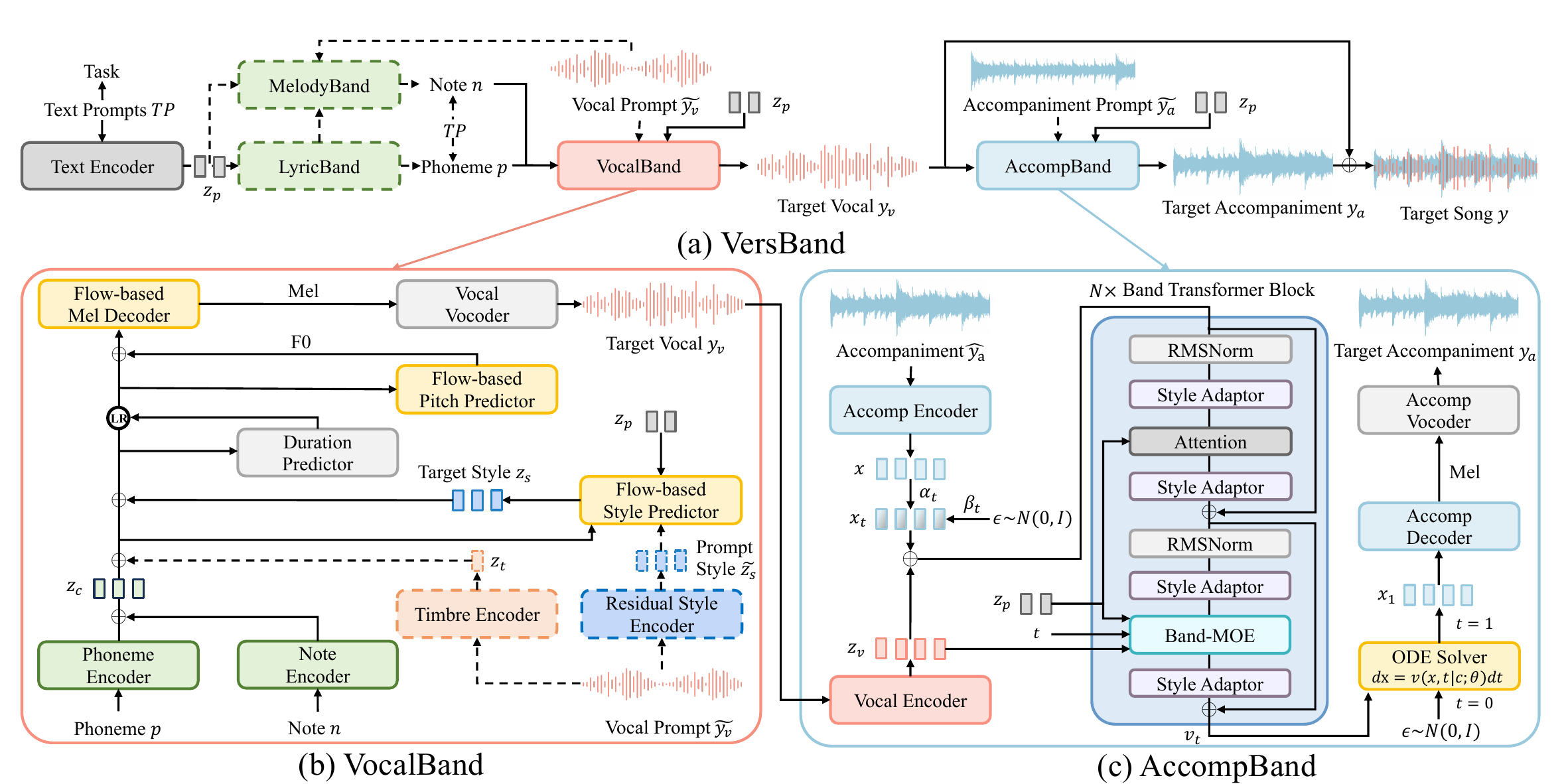}
\caption{The overall architecture of VersBand.
Vocal and accompaniment are generated by VocalBand and AccompBand separately. 
Dashed lines represent optional processes, while LR stands for length regulator. 
}
\label{fig: arch}
\end{figure*}

\subsection{Multi-Task Song Generation}

As shown in Figure \ref{fig: arch} (a), VersBand handles multi-task song generation based on text and audio prompts. 
We design two distinct models, VocalBand for vocals and AccompBand for accompaniments, tailored to their unique characteristics. 
First, we employ a text encoder to generate text tokens $z_p$. 
When lyrics or music scores are not provided, LyricBand and MelodyBand predict phonemes $p$ and notes $n$ (pitch and duration) as target contents.
Next, in Figure \ref{fig: arch} (b), to achieve fast and high-quality vocal generation with granular control, we introduce VocalBand, which decouples the content $z_c$, timbre $z_t$, and style $z_s$. 
Through the Flow-based Pitch Predictor, Mel Decoder, and pre-trained vocoder, the target vocal $y_v$ is synthesized.
Then, in Figure \ref{fig: arch} (c), for the complex nature of accompaniment, we design AccompBand to achieve superior quality, alignment, and control. 
AccompBand uses two encoders to extract embeddings $z_v$ from $y_v$ and $x$ from ground truth (GT) accompaniment $\hat{y_a}$ during training. 
$z_v$ and noise-injected $x_t$ are processed by Band Transformer Blocks with Band-MOE, which selects suitable experts by $z_v$, $z_p$, and time step $t$ for enhanced quality, alignment, and control.
During inference, the ordinary differential equation (ODE) solver, accomp decoder, and vocoder generate the target accompaniment $y_a$. 
$y_v$ and $y_a$ are combined to the final target song $y$.

\subsection{VocalBand}
\label{sec: vocal}

\paragraph{Decomposition}
As shown in Figure \ref{fig: arch}(b), for more personalized and fine-grained control, we disentangle target vocal $y_v$ into content $z_c$, style $z_s$ (e.g., singing methods, emotion, techniques, pronunciation, articulation skills), and timbre $z_t$. 
For $z_c$, phonemes $p$ and notes $n$ (note pitch and duration) are encoded by a phoneme encoder and a note encoder.
Given a vocal prompt $\tilde{y_v}$, the timbre and personalized styles (like pronunciation and articulation skills) should remain consistent. 
We pass $\tilde{y_v}$ through a timbre encoder to obtain $\tilde{z_t}$, while $z_t = \tilde{z_t}$. 
Next, the residual style encoder employs an RQ model \citep{lee2022autoregressive} to extract phoneme-level prompt style $\tilde{z_s}$.
This serves as an information bottleneck to filter out non-style information \citep{zhang2024stylesinger}, ensuring effective decomposition.
The Flow-based Style Predictor uses $z_c$, $z_t$, $\tilde{z_s}$, and text tokens $z_p$ to predict $z_s$, learning both personalized styles of $\tilde{z_s}$ and style control information in $z_p$ (like singing methods).
For more details, please refer to Appendix \ref{app: decom}.

\paragraph{Flow-based Style Predictor}
Singing styles typically exhibit continuous and complex dynamics, involving intricate variations. 
The flow-matching model \citep{liu2022flow} is suitable for generating styles with finer-grained control by modeling styles as a smooth transformation, effectively balancing multiple control inputs, enabling a fast and stable generation of natural and consistent styles.

\begin{figure*}[t]
\centering
\includegraphics[width=1\textwidth]{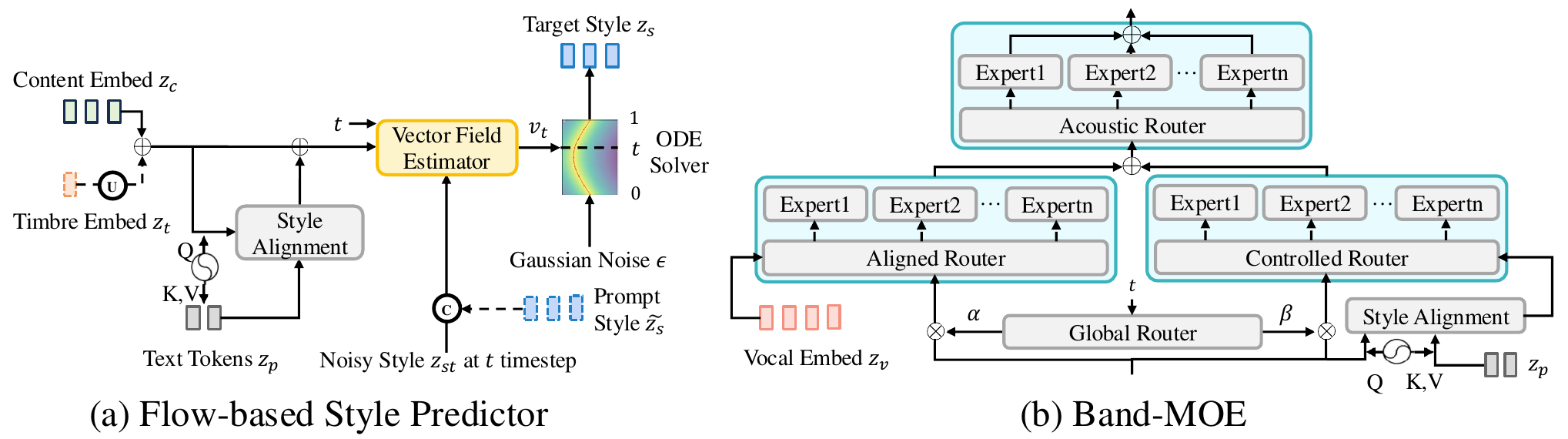}
\caption{The architecture of Flow-based Style Predictor and Band-MOE.
Dashed lines represent optional processes.
C and U represent concatenation and upsampling operations. 
}
\label{fig: arch2}
\end{figure*}

As shown in Figure \ref{fig: arch2} (a), we design the Flow-based Style Predictor using content $z_c$, timbre $z_t$, prompt style $\tilde{z_s}$, and text tokens $z_p$ to predict the target style $z_s$.
With input $z_c$ and $z_t$, we employ a style alignment model with the Scaled Dot-Product Attention mechanism \citep{vaswani2017attention} to align style control information from $z_p$ (e.g., singing methods, emotions, techniques) with contents. 
The fused condition $c$ serves as the condition for an ODE solver, which transforms Gaussian noise $\epsilon$ into $z_s$ along a smooth probability path $p_t(z_{st})$. 

$z_{st}$ is obtained by linear interpolation at time $t$ between $\epsilon$ and $z_s$, which is extracted from the GT vocal by the residual style encoder, thus the target vector field $u(z_{st}, t)=z_s-\epsilon$.
We concatenate $\tilde{z_s}$ with $z_{st}$ to allow $z_s$ to learn personalized styles (e.g., pronunciation, articulation skills). 
The learned vector field $v_t(z_{st}, t | c; \theta)$, predicted by a vector field estimator at each time $t$, ensures smooth interpolation between the initial noise and output, guided by the flow-matching objective, minimizing the distance of learned and true vector fields:
\begin{equation}
\label{equ: loss}
\begin{aligned}
&\mathcal{L}_{style} = \mathbb{E}_{t, p_t(z_{st})} \left\| v_t(z_{st}, t | c; \theta) - (z_s - \epsilon) \right\|^2.
\end{aligned}
\end{equation}
where $p_t(z_{st})$ represents the distribution of $z_{st}$ at time $t$. 
This method ensures the fast and controlled generation of target style $z_s$, learning both personalized styles consistent with $\tilde{z_s}$ and aligned style control information from $z_p$.
Notably, $\tilde{z_s}$ and $z_p$ can be input individually for full style control.
For more details, please refer to Appendix \ref{app: flow} and \ref{app: fsp}.

\paragraph{Flow-based Pitch Predictor and Mel Decoder}
Traditional pitch predictors and mel decoders struggle to capture the dynamic and complex variations in singing voices. 
Thus, we propose the Flow-based Pitch Predictor and Mel Decoder, which use content $z_c$, timbre $z_t$, and style $z_s$ to quickly and robustly generate high-quality F0 and mel-spectrograms with a similar architecture to Flow-based Style Predictor.
Our pitch loss $\mathcal{L}_{pitch}$ and mel loss $\mathcal{L}_{mel}$ are analogous to $\mathcal{L}_{style}$ in Equation \ref{equ: loss}. 
For more details, please refer to Appendix \ref{app: decode}.

\subsection{AccompBand}

\paragraph{Band Transformer Block}
Accompaniment generation is highly complex due to the intricate interplay of various instruments and alignment with vocals, especially for long-sequence generation.
Flow matching enables smooth transformations, leading to stable and quick generation, while transformer models effectively capture intricate long-range dependencies, making flow-based transformers suitable for this task.
As shown in Figure \ref{fig: arch} (b), based on Flag-Dit \citep{gao2024lumina}, we design the Band Transformer Blocks as the vector field estimator.
We add the vocal encoder’s output $z_v$ to the noisy input $x_t$, leveraging the self-attention mechanism for alignment, and use RMSNorm \citep{zhang2019root} and style adaptor with AdaLN \citep{peebles2023scalable} to ensure training stability and style consistency. 
Additionally, we employ rotary positional embeddings (RoPE) \citep{su2024roformer} to capture temporal relationships and the zero-initialized attention mechanism \citep{bachlechner2021rezero} to effectively incorporate conditional information from text tokens $z_p$ into the model.
For more details, please refer to Appendix \ref{app: trans}.

\paragraph{Band-MOE}
To further enhance accompaniment quality, alignment, and control, we propose Band-MOE (Mixture of Experts) in the Band Transformer Block, selecting suitable experts for various conditions.
As shown in Figure \ref{fig: arch2} (d), Band-MOE consists of three expert groups: Aligned MOE, Controlled MOE, and Acoustic MOE, each containing multiple experts. 
Aligned MOE conditions on $z_v$, adjusting inputs to match vocal features like loudness and frequency, selecting suitable experts like one specialized in large loudness and alto range. 
Controlled MOE uses aligned styles in text prompts to select experts for fine-grained style control, such as one for aggressive drums with metal guitar tones. 
Given the varying behavior of the transformer at different noise levels \citep{feng2023ernie}, we design a global router to adjust the weightings for Aligned MOE and Controlled MOE: 
1) at early time steps (near 0), where the hidden representation $h$ is highly noisy, the network prioritizes matching with vocal for coherent reconstruction; 
2) at later time steps (near 1), where $h$ has been largely reconstructed, the network focuses more on refining stylistic details, relying heavily on text prompts. 

Finally, mel-spectrogram patterns exhibit variation across acoustic frequencies \citep{lee2022bigvgan}.
In music, high-frequency components often include the harmonics and overtones of instruments like strings and flutes. 
At the same time, low-frequency content typically encompasses basslines and kick drums providing rhythm and depth.
Since the accomp encoder employs 1D convolutions, the latent should retain the frequency distribution.
Thus, we design Acoustic MOE, selecting experts in different acoustic frequency dimensions for better quality. 
All routing strategies are based on the dense-to-sparse Gumbel-Softmax \citep{nie2021evomoe}, allowing dynamic and efficient expert selection. 
For more details and algorithm, please refer to Appendix \ref{app: moe}.

\paragraph{Classifier-free Guidance}
To further control styles of the generated accompaniment based on input text prompts, we implement the classifier-free guidance (CFG) strategy. 
During AccompBand training, we randomly replace input text tokens $z_p$ with encoded empty strings $\varnothing$ at a probability of 0.2. 
During inference, we modify the output vector field of the Band Transformer blocks as follows:
\begin{equation}
\label{equ: cfg}
\begin{aligned}
& v_{cfg} = \gamma v_t(x, t|z_p;\theta) + (1-\gamma) v_t(x, t|\varnothing ;\theta) ,
\end{aligned}
\end{equation}
where $\gamma$ is the classifier free guidance scale trading off creativity and controllability.
When $\gamma = 1$, $v_{cfg}$ is the same as the original vector field $v_t(x, t|z_p;\theta)$. 
For more details, please refer to Appendix \ref{app: cfg}.

\subsection{Lyric and Melody Generation}

\paragraph{LyricBand}
To enable more personalized tasks, we introduce LyricBand, a system designed to generate complete lyrics based on text prompts. 
Users can design the theme, emotion, and other parameters to generate personalized lyrics. 
We leverage QLoRA \citep{dettmers2024qlora} for efficient fine-tuning of a well-performing open-source bilingual language model Qwen-7B \citep{bai2023qwen}. 
With 4-bit quantization and low-rank adapters, QLoRA enables LyricBand to adapt effectively to lyric generation, enabling customization and creativity.
For more details, please refer to Appendix \ref{app: lyric}.

\paragraph{MelodyBand}
Previous singing voice and song generation models often require users to provide music scores to achieve stable melodies \citep{zhiqing2024text}, lacking customization of the melody. 
Inspired by symbolic music models \citep{dong2018musegan}, we propose MelodyBand, which generates musical notes based on text prompts, lyrics, and vocal prompts. 
We employ a non-autoregressive transformer model to efficiently generate notes. 
After encoding phonemes and timbre, MelodyBand achieves fine-grained melody control by injecting text tokens through cross-attention mechanisms. 
We train MelodyBand with the cross-entropy loss for note pitches and an L2 loss for note durations.
For more details, please refer to Appendix \ref{app: melody}.

\subsection{Training and Inference}

The VocalBand, AccompBand, LyricBand, and MelodyBand are trained separately, and the detailed training details are provided in Appendix \ref{app: train}. 
For inference, our model can accept various prompts for multiple tasks. 
Without input lyrics or music scores, LyricBand and MelodyBand generate phonemes $p$ and notes $n$. 
For song generation or singing style transfer, VocalBand generates the target vocal $y_v$, and AccompBand generates the target accompaniment $y_a$ from Gaussian noise $\epsilon$. 
During music style transfer, AccompBand uses noisy prompt accompaniment $\tilde{y_a}$ instead of $\epsilon$ as input. 
In vocal-to-song, VocalBand is not used, whereas in accompaniment-to-song, notes $n$ extracted from GT accompaniment $\hat{y_a}$ guide VocalBand. 
More inference details can be found in Appendix \ref{app: infer}.

\section{Experiments} 
\label{sec: exp}

\subsection{Experimental Setup}

\paragraph{Dataset}
We train our model using a combination of bilingual web-crawled and open-source song datasets. 
Since there are no publicly available annotated song datasets including vocals and accompaniments, we collect 20k Chinese and English songs from well-known music websites. 
To expand data, we also use open-source singing datasets including GTSinger \citep{zhang2024gtsinger} (30 hours in Chinese and English), M4Singer \citep{zhang2022m4singer} (30 hours in Chinese), and OpenSinger \citep{huang2021multi} (83 hours in Chinese).
After processing and cleaning, we have about 1,000 hours of song data and 1,150 hours of vocal data. 
We also use a filtered subset of LP-MusicCaps-MSD \citep{doh2023lp}, resulting in about 1,200 hours of accompaniment data.
For zero-shot evaluation, we leave out 500 out-of-domain bilingual samples with unseen singers as the test set for each task. 
For more details, please refer to Appendix \ref{app: data}.

\paragraph{Implementation Details}
Mel-spectrograms are driven from raw waveforms with a 24kHz sample rate, 1280 window size, 320 hop size, and 80 mel bins. 
We use 4 layers of Band Transformer Blocks. 
The flow-matching time step is 100 for VocalBand and 1000 for AccompBand during training, while 25 during inference with the Euler ODE solver. 
For more details, please refer to Appendix \ref{app: multi_config}.

\paragraph{Evaluation Metrics}
We conduct both subjective and objective evaluations on generated samples.
For lyric generation, we use overall quality (OVL) and relevance to the prompt (REL) for subjective evaluation.
In melody generation, multiple objective metrics are employed for controllability.
We use the Krumhansl-Schmuckler algorithm to predict the potential key of the generated notes and report the average key accuracy KA. 
We compute the average absolute difference of average pitches (APD) and temporal duration (TD, in seconds).
Then, we employ pitch and duration distribution similarity (PD and DD). 
Melody distance (MD) is also computed using dynamic time warping.

For vocal generation, we conduct MOS (Mean Opinion Score) as the subjective evaluation.
We use MOS-Q for synthesized quality and MOS-C for controllability based on text prompts. 
We also use F0 Frame Error (FFE) as the objective metric.
For singing style transfer, we also employ MOS-S and Cosine Similarity (Cos) to assess singer similarity in timbre and personalized styles of vocal prompts. 

For song generation, raters evaluate audio samples in overall quality (OVL), relevance to the prompt (REL), and alignment with the vocal (ALI). 
For objective evaluation, we calculate Frechet Audio Distance (FAD), Kullback–Leibler Divergence (KLD), and the CLAP score (CLAP). 
Please refer to Appendix \ref{app: eva} for more evaluation details.

\paragraph{Baseline Models}
For lyric generation, we use the original Qwen-7B \citep{bai2023qwen} as the baseline model.
For melody generation, we compare with SongMASS \citep{sheng2021songmass} and MIDI part of MelodyLM \citep{li2024accompanied}. 
For vocal generation, we compare with VISinger2 \citep{zhang2022visinger}, a high-fidelity SVS model, StyleSinger \citep{zhang2024stylesinger}, a zero-shot SVS model, and vocal part of Melodist \citep{zhiqing2024text} and MelodyLM.
For song generation, we compare with Melodist and MelodyLM.
For Melodist and MelodyLM, we use their papers and demos for evaluation, and open-source codes for other models.
Please refer to Appendix \ref{app: base} for more details

\begin{table*}[t]
\small
\centering
\begin{tabular}{l|cccccc}
\toprule
\bfseries{Methods}& KA(\%)$\uparrow$  & APD$\downarrow$   & TD$\downarrow$    & PD(\%)$\uparrow$  & DD(\%)$\uparrow$  & MD $\downarrow$  \\ 
\midrule
SongMASS  & 58.9& 3.78& 2.93 & 55.4& 68.1 & 3.41  \\
MelodyLM & \bf 76.6 & 2.05 & 2.29 &62.8 & 40.8 & 3.62 \\
\midrule
MelodyBand & 72.7 & \bf 1.74 & \bf 1.65 & \bf 65.8 & \bf 70.5 & \bf 3.12 \\
\bottomrule 
\end{tabular}
\caption{Results of melody generation.}
\label{tab: melody}
\end{table*}

\begin{table*}[t]
\centering
\small
\scalebox{1}{
\begin{tabular}{l|ccc|cccc}
\toprule
\multirow{2}{*}{\bfseries{Methods}} & \multicolumn{3}{c|}{\bfseries{Vocal Generation}} & \multicolumn{4}{c}{\bfseries{Singing Style Transfer}}\\
 & {MOS-Q $\uparrow$} & {MOS-C $\uparrow$} & {FFE $\downarrow$}&{MOS-Q $\uparrow$} & {MOS-C $\uparrow$} & {MOS-S $\uparrow$} & {Cos $\uparrow$}\\
\midrule
GT & 4.34 $\pm$ 0.09 & - & - & 4.35 $\pm$ 0.06 & - & -& - \\
\midrule
Melodist   & 3.83$\pm$0.09 & - & 0.12 & - & - & -  & - \\
MelodyLM   & 3.88$\pm$0.10 & - & 0.08 & 3.76$\pm$0.12 &-  & 3.81$\pm$0.12 & - \\
VISinger2  &3.62$\pm$0.07&3.63$\pm$0.09& 0.16 &3.55$\pm$0.11 &3.57$\pm$0.05 & 3.70$\pm$0.08 & 0.82\\
StyleSinger & 3.90$\pm$0.08 & 3.96$\pm$0.05& 0.08 &3.87$\pm$0.06 & 3.86$\pm$0.09 &4.05$\pm$0.05 & 0.89 \\
\midrule
VocalBand  & \bf4.04$\pm$0.08 & \bf 4.02$\pm$0.07& \bf 0.07 & \bf 3.96$\pm$0.10 & \bf 3.95$\pm$0.06 & \bf4.12$\pm$0.04 & \bf0.90\\
\bottomrule
\end{tabular}}
\caption{
Results of vocal generation and singing style transfer.
}
\label{tab: vocal_exp}
\end{table*}

\begin{table}[t]
\small
\centering
\begin{tabular}{l|cc}
\toprule
\bfseries{Methods}             & OVL$\uparrow$   & REL$\uparrow$    \\
\midrule
GT                &  92.31$\pm$1.29               &     84.07$\pm$1.63            \\
\midrule
Qwen-7B         &     74.35$\pm$1.37         &     80.66$\pm$0.92  \\
LyricBand         &     \bf 79.68$\pm$1.05    &   \bf  82.01$\pm$1.13    \\
\bottomrule 
\end{tabular}
\caption{Results of lyric generation.}
\label{tab: lyric}
\end{table}

\subsection{Lyric and Melody Generation}

\paragraph{Lyric Generation}

We evaluate lyric generation models with different text prompts covering aspects such as theme, emotion, genre, style, and specific keywords to generate lyrics.
As shown in Table \ref{tab: lyric}, our fine-tuned LyricBand model outperforms the original Qwen-7B model in overall quality and relevance to text prompts. 
This result highlights the effectiveness of our LyricBand in handling the specific downstream task more proficiently.

\paragraph{Melody Generation}

For MelodyLM, since the melody part is closed-sourced, we directly use the objective metrics reported in the paper. 
As shown in Table \ref{tab: melody}, MelodyBand outperforms SongMASS across all metrics and performs better than MelodyLM except KA. 
Since we use a non-autoregressive transformer architecture, the generation speed is much faster than the autoregressive model of MelodyLM. 
Thus, although MelodyLM has a slightly higher KA, our model is more suitable for the multi-task song generation system.

\subsection{Vocal Generation}

We evaluate VocalBand on both zero-shot vocal generation and singing style transfer tasks using the same test set with unseen singers for fair comparison. 
To enable style control (e.g., singing method, emotion, techniques), we incorporate our text encoder and style alignment models into VISinger2 and StyleSinger. 
Notably, Melodist uses known singer IDs, making it unfair for zero-shot comparisons and incapable of achieving style transfer.
Meanwhile, neither Melodist nor MelodyLM control singing styles, so MOS-C is not provided.

As shown in Table \ref{tab: vocal_exp}, VocalBand consistently outperforms baseline models in both tasks, achieving higher quality (MOS-Q, FFE), similarity (MOS-S, Cos), and controllability (MOS-C). 
This shows the effectiveness of our Flow-based Style Predictor for style control and transfer, as well as the high quality provided by the Flow-based Pitch Predictor and Mel Decoder. 
For more detailed and visualized results, please refer to Appendices \ref{app: svs} and \ref{app: st}.

\subsection{Song Generation}

\begin{table*}[t]
\centering
\small
\scalebox{1}{
\begin{tabular}{l|cccccc}
\toprule
\textbf{Methods}       & {FAD $\downarrow$} & {KLD $\downarrow$} & {CLAP $\uparrow$}  & OVL $\uparrow$ & {REL $\uparrow$} &{ALI $\uparrow$}\\ 
\midrule
Melodist& {3.81}  & {1.34} & {0.39}  &{84.12$\pm$1.54} &{85.97$\pm$1.51} &{74.86$\pm$1.13} \\ 
MelodyLM & 3.42 & 1.35 & 0.35 & 85.23$\pm$1.62 &86.44$\pm$0.90 &75.41$\pm$1.34  \\  
\midrule 
VersBand (w/o lyrics) & 3.37 & 1.30 & 0.50 & 86.65$\pm$0.91 & 85.98$\pm$1.08 & 77.02$\pm$1.33\\
VersBand (w/o scores) & 3.38 & 1.31 & 0.48 & 85.12$\pm$0.77 & 85.15$\pm$1.22 & 75.91$\pm$1.62\\
VersBand (w/o prompts) & 3.55 & 1.35 & -    & 83.49$\pm$1.20 & -              & 74.87$\pm$1.68\\
VersBand (w/ full)  & \bf 3.01 & \bf1.27 & \bf0.58 & \bf87.92$\pm$1.73 & \bf88.03$\pm$0.59 &\bf80.51$\pm$1.66 \\  
\bottomrule
\end{tabular}}
\caption{Results of song generation. }
\label{tab: song}                 
\end{table*}

\begin{table*}[t]
\centering
\small
\scalebox{1}{
\begin{tabular}{l|cccccc}
\toprule
\textbf{Methods}       & {FAD $\downarrow$} & {KLD $\downarrow$} & {CLAP $\uparrow$}  & OVL $\uparrow$ & {REL $\uparrow$} &{ALI $\uparrow$}\\ 
\midrule
VersBand  & 3.01 & 1.27 & 0.58 & 87.92$\pm$1.73 & 88.03$\pm$0.59 & 80.51$\pm$1.66 \\  
\midrule
w/o Band-MOE & 3.29 & 1.34 & 0.42 & 86.11$\pm$1.30  &87.49$\pm$0.84  & 77.59$\pm$1.50\\ 
w/o Aligned MOE & 3.15 & 1.25 & 0.54 & 87.19$\pm$1.14  &88.48$\pm$0.66  & 77.86$\pm$1.35\\  
w/o Controlled MOE & 3.10 & 1.23 & 0.44 & 88.48$\pm$1.74  &87.90$\pm$1.50  & 79.33$\pm$1.63\\
w/o Acoustic MOE & 3.26 & 1.32 & 0.40 & 86.50$\pm$1.55  &87.72$\pm$1.04  & 79.08$\pm$1.24\\
\bottomrule
\end{tabular}}
\caption{Results of ablation study on AccompBand.}
\label{tab: accomp_abl}                 
\end{table*}

\begin{table}
\small
\centering
\begin{tabular}{l|ccc}
\toprule
\bf Methods            & MOS-Q$\uparrow$   & MOS-C$\uparrow$  & FEE$\downarrow$   \\
\midrule
VocalBand           & 4.04$\pm$0.08 & 4.02$\pm$0.07& 0.07    \\
\midrule
w/o Styles           & 3.87$\pm$0.04 & -& 0.09  \\
w/o Pirch Predictor   & 3.79$\pm$0.06 & 3.99$\pm$0.09& 0.09    \\
w/o Mel Decoder       & 3.68$\pm$0.08 & 3.92$\pm$0.07& 0.13  \\
\bottomrule 
\end{tabular}
\caption{Ablation Results of VocalBand.}
\label{tab: vocal_abl}
\end{table}

For song generation evaluation, we remix the generated vocals by VocalBand and accompaniments by AccompBand. 
For MelodyLM and Melodist, we use the objective metrics in their papers and subjectively evaluate the demos on their demo pages.
We test the multi-task capabilities of AccompBand under different conditions: using LyricBand when lyrics are not provided, adding MelodyBand when music scores are missing, using both with no prompts, and finally evaluating full text prompts and optional timbre prompts are provided.
Notably, the REL of MelodyLM and Melodist only considers accompaniment controllability. 
In contrast, for our model, we evaluate lyrics, melody, singing styles, and music styles based on text prompts.

The results are listed in Table \ref{tab: song}, where VersBand demonstrates the highest perceptual quality (FAD, KLD, OVL), the best adherence to text prompts (CLAP, REL), and the most effective alignment (ALI). 
This demonstrates the quality and controllability of VocalBand, as well as the quality, controllability, and alignment of AccompBand.
When some elements in text prompts are removed, VersBand can strike an impressive balance between creativity and stability.
For experiments about more song generation tasks, please refer to Appendix \ref{app: exp}.

\subsection{Ablation Study}

\paragraph{Ablation Study on VocalBand}

We conduct tests on key modules of VocalBand. 
To compare quality, we remove the style information from the Flow-based Style Predictor, and replace the Flow-based Pitch Predictor and Mel Decoder with simpler models from FastSpeech2 \citep{ren2020fastspeech} for comparison. 
As shown in Table \ref{tab: vocal_abl}, we observe that the absence of style representation leads to a decrease in quality, as it cannot generate vocals with rich emotional and stylistic variations, nor can it achieve style control or style transfer.
Additionally, our Flow-based Pitch Predictor and Mel Decoder contribute significantly to the overall quality.

\paragraph{Ablation Study on AccompBand}

We conduct tests on major modules of AccompBand. 
We set the full Band-MOE and three expert groups removed as other baseline models.
As shown in Table \ref{tab: accomp_abl}, removing the Band-MOE results in a decline in all metrics. 
For individual expert groups, we observe that the Aligned MOE affects alignment, while the Controlled MOE impacts controllability. 
The absence of the Acoustic MOE, which handles different acoustic channels, leads to a drop in quality.

\paragraph{Ablation Study on VersBand}

We remove various components from text prompts for evaluation.
As shown in Table \ref{tab: song}, even with a minimum input, VersBand still delivers remarkable performance. 
When listening to songs generated for various tasks on our demo page, it is evident that VersBand shows strong controllability and expressiveness over various text prompts, along with the ability to produce intricate, skillful vocals employing multiple techniques, and complex, well-aligned accompaniments featuring harmonious instrumentation.
\section{Conclusions}

In this paper, we present VersBand, a multi-task song generation framework for synthesizing high-quality, aligned songs with prompt-based control.
We mainly design these key models for the multi-task comprehensive song generation system:
1) VocalBand, a decoupled model leveraging the flow-matching model for singing styles, pitches, and mel-spectrograms generation, allowing fast and high-quality vocal generation with style control.
2) AccompBand, a flow-based transformer model, incorporates the Band-MOE, selecting suitable experts for enhanced quality, alignment, and control.
This model generates controllable, high-quality accompaniments aligned with vocals.
3) Two generation models, LyricBand for lyrics and MelodyBand for melodies, contribute to the versatile song generation system.
Our experimental results demonstrate that VersBand outperforms baseline models across multiple song generation tasks, as measured by both objective and subjective metrics.

\section{Limitations}

In this section, we discuss two main limitations of VersBand and provide potential strategies to address them in future work:
1) 
To achieve comprehensive controllability and high-quality multi-task song generation based on various prompts, VersBand utilizes four sub-models to generate different song components, relying on multiple infrastructures like the flow-based transformer and VAE. 
This results in cumbersome training and inference procedures. 
Future work will explore using a single model to achieve the same multi-task generation capabilities and controllability.
2) 
Our dataset only includes songs in Chinese and English, lacking diversity. 
In the future, we will attempt to build a larger and more comprehensive dataset to enable a wider range of application scenarios.

\section{Ethics Statement}

Large-scale generative models always present ethical challenges. 
VersBand, due to its multi-task song generation capabilities, could potentially be misused for dubbing in entertainment short videos, raising concerns about the infringement of famous singers' copyrights. 
Then, its ability to transfer and control multiple song styles about lyric, melody, singing, and music lowers the requirements for high-quality, personalized, controllable song generation, posing some risks like unfair competition and potential unemployment for professionals in related music and singing occupations.
To mitigate these potential risks, we will explore methods like music watermarking to protect individual privacy.

\section*{Acknowledgements}

This work was supported by National Natural Science Foundation of China under Grant No. 62222211, National Key R\&D Program of China under Grant No.2022ZD0162000, and National Natural Science Foundation of China under Grant No.U24A20326.

\bibliography{custom}

\begin{thebibliography}{56}
\providecommand{\natexlab}[1]{#1}

\bibitem[{Achiam et~al.(2023)Achiam, Adler, Agarwal, Ahmad, Akkaya, Aleman, Almeida, Altenschmidt, Altman, Anadkat et~al.}]{achiam2023gpt}
Josh Achiam, Steven Adler, Sandhini Agarwal, Lama Ahmad, Ilge Akkaya, Florencia~Leoni Aleman, Diogo Almeida, Janko Altenschmidt, Sam Altman, Shyamal Anadkat, et~al. 2023.
\newblock Gpt-4 technical report.
\newblock \emph{arXiv preprint arXiv:2303.08774}.

\bibitem[{Agostinelli et~al.(2023)Agostinelli, Denk, Borsos, Engel, Verzetti, Caillon, Huang, Jansen, Roberts, Tagliasacchi et~al.}]{agostinelli2023musiclm}
Andrea Agostinelli, Timo~I Denk, Zal{\'a}n Borsos, Jesse Engel, Mauro Verzetti, Antoine Caillon, Qingqing Huang, Aren Jansen, Adam Roberts, Marco Tagliasacchi, et~al. 2023.
\newblock Musiclm: Generating music from text.
\newblock \emph{arXiv preprint arXiv:2301.11325}.

\bibitem[{Bachlechner et~al.(2021)Bachlechner, Majumder, Mao, Cottrell, and McAuley}]{bachlechner2021rezero}
Thomas Bachlechner, Bodhisattwa~Prasad Majumder, Henry Mao, Gary Cottrell, and Julian McAuley. 2021.
\newblock Rezero is all you need: Fast convergence at large depth.
\newblock In \emph{Uncertainty in Artificial Intelligence}, pages 1352--1361. PMLR.

\bibitem[{Bai et~al.(2023)Bai, Bai, Chu, Cui, Dang, Deng, Fan, Ge, Han, Huang et~al.}]{bai2023qwen}
Jinze Bai, Shuai Bai, Yunfei Chu, Zeyu Cui, Kai Dang, Xiaodong Deng, Yang Fan, Wenbin Ge, Yu~Han, Fei Huang, et~al. 2023.
\newblock Qwen technical report.
\newblock \emph{arXiv preprint arXiv:2309.16609}.

\bibitem[{Bain et~al.(2023)Bain, Huh, Han, and Zisserman}]{bain2022whisperx}
Max Bain, Jaesung Huh, Tengda Han, and Andrew Zisserman. 2023.
\newblock Whisperx: Time-accurate speech transcription of long-form audio.
\newblock \emph{INTERSPEECH 2023}.

\bibitem[{Berndt and Clifford(1994)}]{berndt1994using}
Donald~J Berndt and James Clifford. 1994.
\newblock Using dynamic time warping to find patterns in time series.
\newblock In \emph{KDD workshop}, volume~10, pages 359--370. Seattle, WA, USA:.

\bibitem[{Chen et~al.(2020)Chen, Tan, Luan, Qin, and Liu}]{chen2020hifisinger}
Jiawei Chen, Xu~Tan, Jian Luan, Tao Qin, and Tie-Yan Liu. 2020.
\newblock Hifisinger: Towards high-fidelity neural singing voice synthesis.
\newblock \emph{arXiv preprint arXiv:2009.01776}.

\bibitem[{Chen et~al.(2024)Chen, Wu, Liu, Nezhurina, Berg-Kirkpatrick, and Dubnov}]{chen2024musicldm}
Ke~Chen, Yusong Wu, Haohe Liu, Marianna Nezhurina, Taylor Berg-Kirkpatrick, and Shlomo Dubnov. 2024.
\newblock Musicldm: Enhancing novelty in text-to-music generation using beat-synchronous mixup strategies.
\newblock In \emph{ICASSP 2024-2024 IEEE International Conference on Acoustics, Speech and Signal Processing (ICASSP)}, pages 1206--1210. IEEE.

\bibitem[{Choi and Nam(2022)}]{choi2022melody}
Soonbeom Choi and Juhan Nam. 2022.
\newblock A melody-unsupervision model for singing voice synthesis.
\newblock In \emph{ICASSP 2022-2022 IEEE International Conference on Acoustics, Speech and Signal Processing (ICASSP)}, pages 7242--7246. IEEE.

\bibitem[{Chung et~al.(2024)Chung, Hou, Longpre, Zoph, Tay, Fedus, Li, Wang, Dehghani, Brahma et~al.}]{chung2024scaling}
Hyung~Won Chung, Le~Hou, Shayne Longpre, Barret Zoph, Yi~Tay, William Fedus, Yunxuan Li, Xuezhi Wang, Mostafa Dehghani, Siddhartha Brahma, et~al. 2024.
\newblock Scaling instruction-finetuned language models.
\newblock \emph{Journal of Machine Learning Research}, 25(70):1--53.

\bibitem[{Copet et~al.(2024)Copet, Kreuk, Gat, Remez, Kant, Synnaeve, Adi, and D{\'e}fossez}]{copet2024simple}
Jade Copet, Felix Kreuk, Itai Gat, Tal Remez, David Kant, Gabriel Synnaeve, Yossi Adi, and Alexandre D{\'e}fossez. 2024.
\newblock Simple and controllable music generation.
\newblock \emph{Advances in Neural Information Processing Systems}, 36.

\bibitem[{Dettmers et~al.(2024)Dettmers, Pagnoni, Holtzman, and Zettlemoyer}]{dettmers2024qlora}
Tim Dettmers, Artidoro Pagnoni, Ari Holtzman, and Luke Zettlemoyer. 2024.
\newblock Qlora: Efficient finetuning of quantized llms.
\newblock \emph{Advances in Neural Information Processing Systems}, 36.

\bibitem[{Doh et~al.(2023)Doh, Choi, Lee, and Nam}]{doh2023lp}
SeungHeon Doh, Keunwoo Choi, Jongpil Lee, and Juhan Nam. 2023.
\newblock Lp-musiccaps: Llm-based pseudo music captioning.
\newblock \emph{arXiv preprint arXiv:2307.16372}.

\bibitem[{Dong et~al.(2018)Dong, Hsiao, Yang, and Yang}]{dong2018musegan}
Hao-Wen Dong, Wen-Yi Hsiao, Li-Chia Yang, and Yi-Hsuan Yang. 2018.
\newblock Musegan: Multi-track sequential generative adversarial networks for symbolic music generation and accompaniment.
\newblock In \emph{Proceedings of the AAAI Conference on Artificial Intelligence}, volume~32.

\bibitem[{Fedus et~al.(2022)Fedus, Zoph, and Shazeer}]{fedus2022switch}
William Fedus, Barret Zoph, and Noam Shazeer. 2022.
\newblock Switch transformers: Scaling to trillion parameter models with simple and efficient sparsity.
\newblock \emph{Journal of Machine Learning Research}, 23(120):1--39.

\bibitem[{Feng et~al.(2023)Feng, Zhang, Yu, Fang, Li, Chen, Lu, Liu, Yin, Feng et~al.}]{feng2023ernie}
Zhida Feng, Zhenyu Zhang, Xintong Yu, Yewei Fang, Lanxin Li, Xuyi Chen, Yuxiang Lu, Jiaxiang Liu, Weichong Yin, Shikun Feng, et~al. 2023.
\newblock Ernie-vilg 2.0: Improving text-to-image diffusion model with knowledge-enhanced mixture-of-denoising-experts.
\newblock In \emph{Proceedings of the IEEE/CVF Conference on Computer Vision and Pattern Recognition}, pages 10135--10145.

\bibitem[{Gao et~al.(2024)Gao, Zhuo, Lin, Liu, Chen, Du, Xie, Luo, Qiu, Zhang et~al.}]{gao2024lumina}
Peng Gao, Le~Zhuo, Ziyi Lin, Chris Liu, Junsong Chen, Ruoyi Du, Enze Xie, Xu~Luo, Longtian Qiu, Yuhang Zhang, et~al. 2024.
\newblock Lumina-t2x: Transforming text into any modality, resolution, and duration via flow-based large diffusion transformers.
\newblock \emph{arXiv preprint arXiv:2405.05945}.

\bibitem[{Guo et~al.(2025)Guo, Zhang, Pan, Huang, Tang, Li, Hong, Wang, and Zhao}]{guo2025techsinger}
Wenxiang Guo, Yu~Zhang, Changhao Pan, Rongjie Huang, Li~Tang, Ruiqi Li, Zhiqing Hong, Yongqi Wang, and Zhou Zhao. 2025.
\newblock Techsinger: Technique controllable multilingual singing voice synthesis via flow matching.
\newblock \emph{arXiv preprint arXiv:2502.12572}.

\bibitem[{Huang et~al.(2022)Huang, Jansen, Lee, Ganti, Li, and Ellis}]{huang2022mulan}
Qingqing Huang, Aren Jansen, Joonseok Lee, Ravi Ganti, Judith~Yue Li, and Daniel~PW Ellis. 2022.
\newblock Mulan: A joint embedding of music audio and natural language.
\newblock \emph{arXiv preprint arXiv:2208.12415}.

\bibitem[{Huang et~al.(2021)Huang, Chen, Ren, Liu, Cui, and Zhao}]{huang2021multi}
Rongjie Huang, Feiyang Chen, Yi~Ren, Jinglin Liu, Chenye Cui, and Zhou Zhao. 2021.
\newblock Multi-singer: Fast multi-singer singing voice vocoder with a large-scale corpus.
\newblock In \emph{Proceedings of the 29th ACM International Conference on Multimedia}, pages 3945--3954.

\bibitem[{Jang et~al.(2016)Jang, Gu, and Poole}]{jang2016categorical}
Eric Jang, Shixiang Gu, and Ben Poole. 2016.
\newblock Categorical reparameterization with gumbel-softmax.
\newblock \emph{arXiv preprint arXiv:1611.01144}.

\bibitem[{Jiang et~al.(2024)Jiang, Liu, Ren, He, Ye, Ji, Yang, Zhang, Wei, Wang et~al.}]{jiang2024mega}
Ziyue Jiang, Jinglin Liu, Yi~Ren, Jinzheng He, Zhenhui Ye, Shengpeng Ji, Qian Yang, Chen Zhang, Pengfei Wei, Chunfeng Wang, et~al. 2024.
\newblock Mega-tts 2: Boosting prompting mechanisms for zero-shot speech synthesis.
\newblock In \emph{The Twelfth International Conference on Learning Representations}.

\bibitem[{Jiang et~al.(2025)Jiang, Ren, Li, Ji, Ye, Zhang, Jionghao, Yang, Zuo, Zhang, Liu, Yin, and Zhao}]{jiang2025sparse}
Ziyue Jiang, Yi~Ren, Ruiqi Li, Shengpeng Ji, Zhenhui Ye, Chen Zhang, Bai Jionghao, Xiaoda Yang, Jialong Zuo, Yu~Zhang, Rui Liu, Xiang Yin, and Zhou Zhao. 2025.
\newblock Sparse alignment enhanced latent diffusion transformer for zero-shot speech synthesis.
\newblock \emph{arXiv preprint arXiv:2502.18924}.

\bibitem[{Kilgour et~al.(2018)Kilgour, Zuluaga, Roblek, and Sharifi}]{kilgour2018fr}
Kevin Kilgour, Mauricio Zuluaga, Dominik Roblek, and Matthew Sharifi. 2018.
\newblock Fr$\backslash$'echet audio distance: A metric for evaluating music enhancement algorithms.
\newblock \emph{arXiv preprint arXiv:1812.08466}.

\bibitem[{Kim et~al.(2024)Kim, Kang, and Lee}]{kim2024adversarial}
Tae-Woo Kim, Min-Su Kang, and Gyeong-Hoon Lee. 2024.
\newblock \href {https://arxiv.org/abs/2206.11558} {Adversarial multi-task learning for disentangling timbre and pitch in singing voice synthesis}.
\newblock \emph{Preprint}, arXiv:2206.11558.

\bibitem[{Kingma and Welling(2013)}]{kingma2013auto}
Diederik~P Kingma and Max Welling. 2013.
\newblock Auto-encoding variational bayes.
\newblock \emph{arXiv preprint arXiv:1312.6114}.

\bibitem[{Kong et~al.(2020)Kong, Kim, and Bae}]{kong2020hifi}
Jungil Kong, Jaehyeon Kim, and Jaekyoung Bae. 2020.
\newblock Hifi-gan: Generative adversarial networks for efficient and high fidelity speech synthesis.
\newblock \emph{Advances in neural information processing systems}, 33:17022--17033.

\bibitem[{Lee et~al.(2022{\natexlab{a}})Lee, Kim, Kim, Cho, and Han}]{lee2022autoregressive}
Doyup Lee, Chiheon Kim, Saehoon Kim, Minsu Cho, and Wook-Shin Han. 2022{\natexlab{a}}.
\newblock Autoregressive image generation using residual quantization.
\newblock In \emph{Proceedings of the IEEE/CVF Conference on Computer Vision and Pattern Recognition}, pages 11523--11532.

\bibitem[{Lee et~al.(2022{\natexlab{b}})Lee, Ping, Ginsburg, Catanzaro, and Yoon}]{lee2022bigvgan}
Sang-gil Lee, Wei Ping, Boris Ginsburg, Bryan Catanzaro, and Sungroh Yoon. 2022{\natexlab{b}}.
\newblock Bigvgan: A universal neural vocoder with large-scale training.
\newblock \emph{arXiv preprint arXiv:2206.04658}.

\bibitem[{Lei et~al.(2024)Lei, Zhou, Tang, Lam, Liu, Liu, Wu, Kang, Wu, and Meng}]{lei2024songcreator}
Shun Lei, Yixuan Zhou, Boshi Tang, Max~WY Lam, Feng Liu, Hangyu Liu, Jingcheng Wu, Shiyin Kang, Zhiyong Wu, and Helen Meng. 2024.
\newblock Songcreator: Lyrics-based universal song generation.
\newblock \emph{arXiv preprint arXiv:2409.06029}.

\bibitem[{Li et~al.(2024{\natexlab{a}})Li, Hong, Wang, Zhang, Huang, Zheng, and Zhao}]{li2024accompanied}
Ruiqi Li, Zhiqing Hong, Yongqi Wang, Lichao Zhang, Rongjie Huang, Siqi Zheng, and Zhou Zhao. 2024{\natexlab{a}}.
\newblock Accompanied singing voice synthesis with fully text-controlled melody.
\newblock \emph{arXiv preprint arXiv:2407.02049}.

\bibitem[{Li et~al.(2024{\natexlab{b}})Li, Zhang, Wang, Hong, Huang, and Zhao}]{li2024robust}
Ruiqi Li, Yu~Zhang, Yongqi Wang, Zhiqing Hong, Rongjie Huang, and Zhou Zhao. 2024{\natexlab{b}}.
\newblock \href {https://arxiv.org/abs/2405.09940} {Robust singing voice transcription serves synthesis}.
\newblock \emph{Preprint}, arXiv:2405.09940.

\bibitem[{Lipman et~al.(2022)Lipman, Chen, Ben-Hamu, Nickel, and Le}]{lipman2022flow}
Yaron Lipman, Ricky~TQ Chen, Heli Ben-Hamu, Maximilian Nickel, and Matthew Le. 2022.
\newblock Flow matching for generative modeling.
\newblock In \emph{The Eleventh International Conference on Learning Representations}.

\bibitem[{Liu et~al.(2022)Liu, Gong et~al.}]{liu2022flow}
Xingchao Liu, Chengyue Gong, et~al. 2022.
\newblock Flow straight and fast: Learning to generate and transfer data with rectified flow.
\newblock In \emph{The Eleventh International Conference on Learning Representations}.

\bibitem[{Mao et~al.(2017)Mao, Li, Xie, Lau, Wang, and Paul~Smolley}]{mao2017least}
Xudong Mao, Qing Li, Haoran Xie, Raymond~YK Lau, Zhen Wang, and Stephen Paul~Smolley. 2017.
\newblock Least squares generative adversarial networks.
\newblock In \emph{Proceedings of the IEEE international conference on computer vision}, pages 2794--2802.

\bibitem[{McAuliffe et~al.(2017)McAuliffe, Socolof, Mihuc, Wagner, and Sonderegger}]{mcauliffe2017montreal}
Michael McAuliffe, Michaela Socolof, Sarah Mihuc, Michael Wagner, and Morgan Sonderegger. 2017.
\newblock Montreal forced aligner: Trainable text-speech alignment using kaldi.
\newblock In \emph{Interspeech}, volume 2017, pages 498--502.

\bibitem[{Nie et~al.(2021)Nie, Miao, Cao, Ma, Liu, Xue, Miao, Liu, Yang, and Cui}]{nie2021evomoe}
Xiaonan Nie, Xupeng Miao, Shijie Cao, Lingxiao Ma, Qibin Liu, Jilong Xue, Youshan Miao, Yi~Liu, Zhi Yang, and Bin Cui. 2021.
\newblock Evomoe: An evolutional mixture-of-experts training framework via dense-to-sparse gate.
\newblock \emph{arXiv preprint arXiv:2112.14397}.

\bibitem[{Peebles and Xie(2023)}]{peebles2023scalable}
William Peebles and Saining Xie. 2023.
\newblock Scalable diffusion models with transformers.
\newblock In \emph{Proceedings of the IEEE/CVF International Conference on Computer Vision}, pages 4195--4205.

\bibitem[{Qian et~al.(2019)Qian, Zhang, Chang, Yang, and Hasegawa-Johnson}]{qian2019autovc}
Kaizhi Qian, Yang Zhang, Shiyu Chang, Xuesong Yang, and Mark Hasegawa-Johnson. 2019.
\newblock Autovc: Zero-shot voice style transfer with only autoencoder loss.
\newblock In \emph{International Conference on Machine Learning}, pages 5210--5219. PMLR.

\bibitem[{Ren et~al.(2020)Ren, Hu, Tan, Qin, Zhao, Zhao, and Liu}]{ren2020fastspeech}
Yi~Ren, Chenxu Hu, Xu~Tan, Tao Qin, Sheng Zhao, Zhou Zhao, and Tie-Yan Liu. 2020.
\newblock Fastspeech 2: Fast and high-quality end-to-end text to speech.
\newblock \emph{arXiv preprint arXiv:2006.04558}.

\bibitem[{Sheng et~al.(2021)Sheng, Song, Tan, Ren, Ye, Zhang, and Qin}]{sheng2021songmass}
Zhonghao Sheng, Kaitao Song, Xu~Tan, Yi~Ren, Wei Ye, Shikun Zhang, and Tao Qin. 2021.
\newblock Songmass: Automatic song writing with pre-training and alignment constraint.
\newblock In \emph{Proceedings of the AAAI Conference on Artificial Intelligence}, volume~35, pages 13798--13805.

\bibitem[{Su et~al.(2024)Su, Ahmed, Lu, Pan, Bo, and Liu}]{su2024roformer}
Jianlin Su, Murtadha Ahmed, Yu~Lu, Shengfeng Pan, Wen Bo, and Yunfeng Liu. 2024.
\newblock Roformer: Enhanced transformer with rotary position embedding.
\newblock \emph{Neurocomputing}, 568:127063.

\bibitem[{Team(2021)}]{team2021silero}
Silero Team. 2021.
\newblock Silero vad: pre-trained enterprise-grade voice activity detector (vad), number detector and language classifier.
\newblock \emph{Retrieved March}, 31:2023.

\bibitem[{Van Den~Oord et~al.(2016)Van Den~Oord, Dieleman, Zen, Simonyan, Vinyals, Graves, Kalchbrenner, Senior, Kavukcuoglu et~al.}]{van2016wavenet}
Aaron Van Den~Oord, Sander Dieleman, Heiga Zen, Karen Simonyan, Oriol Vinyals, Alex Graves, Nal Kalchbrenner, Andrew Senior, Koray Kavukcuoglu, et~al. 2016.
\newblock Wavenet: A generative model for raw audio.
\newblock \emph{arXiv preprint arXiv:1609.03499}, 12.

\bibitem[{Vaswani et~al.(2017)Vaswani, Shazeer, Parmar, Uszkoreit, Jones, Gomez, Kaiser, and Polosukhin}]{vaswani2017attention}
Ashish Vaswani, Noam Shazeer, Niki Parmar, Jakob Uszkoreit, Llion Jones, Aidan~N Gomez, {\L}ukasz Kaiser, and Illia Polosukhin. 2017.
\newblock Attention is all you need.
\newblock In \emph{Advances in neural information processing systems}, pages 5998--6008.

\bibitem[{Wang et~al.(2024)Wang, Hu, Huang, Hong, Li, Liu, You, Jin, and Zhao}]{wang2024prompt}
Yongqi Wang, Ruofan Hu, Rongjie Huang, Zhiqing Hong, Ruiqi Li, Wenrui Liu, Fuming You, Tao Jin, and Zhou Zhao. 2024.
\newblock Prompt-singer: Controllable singing-voice-synthesis with natural language prompt.
\newblock \emph{arXiv preprint arXiv:2403.11780}.

\bibitem[{Wei et~al.(2023)Wei, Cao, Dan, and Chen}]{wei2023rmvpe}
Haojie Wei, Xueke Cao, Tangpeng Dan, and Yueguo Chen. 2023.
\newblock Rmvpe: A robust model for vocal pitch estimation in polyphonic music.
\newblock \emph{arXiv preprint arXiv:2306.15412}.

\bibitem[{Wu et~al.(2023)Wu, Chen, Zhang, Hui, Berg-Kirkpatrick, and Dubnov}]{wu2023large}
Yusong Wu, Ke~Chen, Tianyu Zhang, Yuchen Hui, Taylor Berg-Kirkpatrick, and Shlomo Dubnov. 2023.
\newblock Large-scale contrastive language-audio pretraining with feature fusion and keyword-to-caption augmentation.
\newblock In \emph{ICASSP 2023-2023 IEEE International Conference on Acoustics, Speech and Signal Processing (ICASSP)}, pages 1--5. IEEE.

\bibitem[{Yu et~al.(2021)Yu, Li, Koh, Zhang, Pang, Qin, Ku, Xu, Baldridge, and Wu}]{yu2021vector}
Jiahui Yu, Xin Li, Jing~Yu Koh, Han Zhang, Ruoming Pang, James Qin, Alexander Ku, Yuanzhong Xu, Jason Baldridge, and Yonghui Wu. 2021.
\newblock Vector-quantized image modeling with improved vqgan.
\newblock \emph{arXiv preprint arXiv:2110.04627}.

\bibitem[{Zhang and Sennrich(2019)}]{zhang2019root}
Biao Zhang and Rico Sennrich. 2019.
\newblock Root mean square layer normalization.
\newblock \emph{Advances in Neural Information Processing Systems}, 32.

\bibitem[{Zhang et~al.(2022{\natexlab{a}})Zhang, Li, Wang, Deng, Liu, Ren, He, Huang, Zhu, Chen et~al.}]{zhang2022m4singer}
Lichao Zhang, Ruiqi Li, Shoutong Wang, Liqun Deng, Jinglin Liu, Yi~Ren, Jinzheng He, Rongjie Huang, Jieming Zhu, Xiao Chen, et~al. 2022{\natexlab{a}}.
\newblock M4singer: A multi-style, multi-singer and musical score provided mandarin singing corpus.
\newblock \emph{Advances in Neural Information Processing Systems}, 35:6914--6926.

\bibitem[{Zhang et~al.(2022{\natexlab{b}})Zhang, Xue, Li, Xie, Guo, Zhang, and Gong}]{zhang2022visinger}
Yongmao Zhang, Heyang Xue, Hanzhao Li, Lei Xie, Tingwei Guo, Ruixiong Zhang, and Caixia Gong. 2022{\natexlab{b}}.
\newblock \href {https://arxiv.org/abs/2211.02903} {Visinger 2: High-fidelity end-to-end singing voice synthesis enhanced by digital signal processing synthesizer}.
\newblock \emph{Preprint}, arXiv:2211.02903.

\bibitem[{Zhang et~al.(2024{\natexlab{a}})Zhang, Huang, Li, He, Xia, Chen, Duan, Huai, and Zhao}]{zhang2024stylesinger}
Yu~Zhang, Rongjie Huang, Ruiqi Li, JinZheng He, Yan Xia, Feiyang Chen, Xinyu Duan, Baoxing Huai, and Zhou Zhao. 2024{\natexlab{a}}.
\newblock Stylesinger: Style transfer for out-of-domain singing voice synthesis.
\newblock In \emph{Proceedings of the AAAI Conference on Artificial Intelligence}, volume~38, pages 19597--19605.

\bibitem[{Zhang et~al.(2024{\natexlab{b}})Zhang, Jiang, Li, Pan, He, Huang, Wang, and Zhao}]{zhang2024tcsinger}
Yu~Zhang, Ziyue Jiang, Ruiqi Li, Changhao Pan, Jinzheng He, Rongjie Huang, Chuxin Wang, and Zhou Zhao. 2024{\natexlab{b}}.
\newblock Tcsinger: Zero-shot singing voice synthesis with style transfer and multi-level style control.
\newblock \emph{arXiv preprint arXiv:2409.15977}.

\bibitem[{Zhang et~al.(2024{\natexlab{c}})Zhang, Pan, Guo, Li, Zhu, Wang, Xu, Lu, Hong, Wang et~al.}]{zhang2024gtsinger}
Yu~Zhang, Changhao Pan, Wenxiang Guo, Ruiqi Li, Zhiyuan Zhu, Jialei Wang, Wenhao Xu, Jingyu Lu, Zhiqing Hong, Chuxin Wang, et~al. 2024{\natexlab{c}}.
\newblock Gtsinger: A global multi-technique singing corpus with realistic music scores for all singing tasks.
\newblock \emph{arXiv preprint arXiv:2409.13832}.

\bibitem[{Zhiqing et~al.(2024)Zhiqing, Rongjie, Xize, Yongqi, Ruiqi, Fuming, Zhou, and Zhimeng}]{zhiqing2024text}
Hong Zhiqing, Huang Rongjie, Cheng Xize, Wang Yongqi, Li~Ruiqi, You Fuming, Zhao Zhou, and Zhang Zhimeng. 2024.
\newblock Text-to-song: Towards controllable music generation incorporating vocals and accompaniment.
\newblock \emph{arXiv preprint arXiv:2404.09313}.

\end{thebibliography}

\newpage
\appendix

\section{Rectified flow-matching}
\label{app: flow}

In this section, we introduce the flow-matching generative method, as described by \citet{liu2022flow}. 
In generative modeling, the true data distribution is denoted as $q(x_1)$, which can be sampled but lacks an accessible density function. 
Consider a probability path $p_t(x_t)$, where $x_0 \sim p_0(x)$ represents a known simple distribution (e.g., a standard Gaussian), and $x_1 \sim p_1(x)$ approximates the real data distribution. 
The objective of flow-matching is to model this probability path directly, expressed as an ordinary differential equation (ODE):
\begin{equation}
\begin{aligned}
&\mathrm{d}x = u(x, t) \mathrm{d}t, \quad t \in [0, 1],
\end{aligned}
\end{equation}
where $u(x, t)$ denotes the target vector field, and $t$ is the time index. 
If the vector field $u$ is known, realistic data can be recovered by reversing the flow. 
To approximate $u$, a vector field estimator $v(\cdot)$ is used, with the flow-matching objective defined as:
\begin{equation}
\begin{aligned}
&\mathcal{L}_{\mathrm{FM}}(\theta) = \mathbb{E}_{t, p_t(x)} \left\| v(x, t; \theta) - u(x, t) \right\|^2,
\end{aligned}
\end{equation}
where $p_t(x)$ denotes the distribution of $x$ at time $t$. 
To enable conditional generation, we add conditional information $c$, leading to the conditional flow-matching objective \citep{lipman2022flow}:
\begin{equation}
\begin{aligned}
&\mathcal{L}_{\mathrm{CFM}}(\theta) = \\
& \mathbb{E}_{t, p_1(x_1), p_t(x | x_1)} \left\| v(x, t | c; \theta) - u(x, t | x_1, c) \right\|^2.
\end{aligned}
\end{equation}
Flow-matching proposes a straight path from noise to data. 
Specifically, we use a linear interpolation between the data $x_1$ and Gaussian noise $x_0$ to generate samples at time $t$: 
\begin{equation}
\begin{aligned}
&x_t = (1 - t) x_0 + t x_1.
\end{aligned}
\end{equation}
Thus, the conditional vector field becomes $u(x, t | x_1, c) = x_1 - x_0$, and the rectified flow-matching (RFM) loss used for gradient descent is:
\begin{equation}
\label{equ: flow}
\begin{aligned}
&\left\| v(x, t | c; \theta) - (x_1 - x_0) \right\|^2.
\end{aligned}
\end{equation}
If the vector field $u$ is estimated correctly, we can generate realistic data by propagating Gaussian noise through an ODE solver at discrete time steps. 
A widely used method for solving the reverse flow is the Euler ODE:
\begin{equation}
\begin{aligned}
&x_{t + \epsilon} = x + \epsilon v(x, t | c; \theta),
\label{eq:euler}
\end{aligned}
\end{equation}
where $\epsilon$ is the step size. 
In our VocalBand, we use content, timbre, prompt style, text tokens, and other inputs for each task as conditioning information $c$, while the target data $x_1$ consists of target style, F0, or mel-spectrograms.
In our AccompBand, we use timestep, text tokens, and vocal embedding as conditioning information $c$, while the target data $x_1$ is the accompaniment embedding.

Moreover, flow matching models require 100 to 1000 steps during training, but since they generate a straight path, they only require 25 or fewer steps during inference, making the generation highly efficient for fast generation. 
Additionally, flow-matching models ensure stable and high-quality generation due to their ability to model smooth transitions between noise and data, maintaining fidelity throughout the process. 
This stability is crucial for complex generation tasks, as it reduces artifacts and enhances the consistency of the output across various conditions.

\section{VersBand Details}
\label{app: mod}

\subsection{Model Details}
\label{app: multi_config}

Our VersBand framework consists of four models: VocalBand, AccompBand, LyricBand, and MelodyBand. 
For the text encoder, we use FLAN-T5-large \citep{chung2024scaling}. 
Our vocoder is the pre-trained HiFi-GAN \citep{kong2020hifi}. 
For detailed hyperparameters of each component, please refer to Appendix \ref{app: vocal_config}, \ref{app: accomp_config}, \ref{app: lyric}, and \ref{app: melody}.

For training details, we set the sample rate to 24kHz, the window size to 1280, the hop size to 320, and the number of mel bins to 80 to derive mel-spectrograms from raw waveforms.
We train VocalBand on 4 NVIDIA RTX-4090 GPUs for 200k steps. 
The Adam optimizer is used with $\beta_1 = 0.9$ and $\beta_2 = 0.98$. 
AccompBand is trained on 8 NVIDIA RTX-4090 GPUs for 80k steps, using the AdamW optimizer with a base learning rate of $3 \times 10^{-6}$. 
The pre-trained accomp encoder and decoder are trained on 4 NVIDIA RTX-4090 GPUs for 40k steps.
MelodyBand is trained for 30k steps until convergence on 4 NVIDIA RTX-4090 GPUs. 
LyricBand is fine-tuned for 4k steps until convergence on 4 NVIDIA RTX-4090 GPUs.

\subsection{Training Procedures}
\label{app: train}

For VocalBand, the final loss terms in the training phase include the following components: 
1) $\mathcal{L}_{commit}$: the commitment loss for the residual style encoder in Equation \ref{equ: commit}; 
2) $\mathcal{L}_{style}$: the flow matching loss of Flow-based Style Predictor in Equation \ref{equ: loss};
3) $\mathcal{L}_{pitch}$: the flow matching loss of Flow-based Pitch Predictor;
4) $\mathcal{L}_{mel}$: the flow matching loss of Flow-based Mel Decoder;
5) $\mathcal{L}_{dur}$: the MSE duration loss between the predicted and the GT phoneme-level duration in the log scale.

As for AccompBand, the final loss terms during training consist of the following aspects:
1) $\mathcal{L}_{balance}$: the load-balancing loss for each expert group in Band-MOE in Equation \ref{equ: balance};
2) $\mathcal{L}_{flow}$: the flow matching loss of AccompBand.

For the pre-trained accomp encoder and accomp decoder, the final loss terms include:
1) $\mathcal{L}_{rec}$: the L2 reconstruction loss;
2) $\mathcal{L}_{adv}$: the LSGAN-styled adversarial loss in GAN discriminator.

Regarding MelodyBand, the final loss terms for training involve:
1) $\mathcal{L}_{pitch}$: the cross-entropy loss for note pitches in Equation \ref{equ: pitch};
2) $\mathcal{L}_{duration}$: the L2 loss for note durations in Equation \ref{equ: dur}.

\subsection{Multi-Task Inference Procedures}
\label{app: infer}

During inference, we can achieve multiple song generation tasks based on text and audio prompts.
If full lyrics are not provided, LyricBand generates phonemes $p$ based on the text tokens $z_p$. Without input music scores, MelodyBand generates notes $n$ (note pitches and note durations) based on lyrics, text prompts, and optional vocal prompts.

For the song generation task, VocalBand generates the target vocal $y_v$ based on $n$ and $p$ as contents, along with $z_p$ to control styles, timbre prompt $z_t$ is optional. 
AccompBand generates the target accompaniment $y_a$ from Gaussian noise $\epsilon$ and $y_v$.

To conduct singing style transfer, VocalBand additionally takes a vocal prompt $\tilde{y_a}$ as input to extract timbre $z_t$ and prompt style $\tilde{z_s}$. 
The target vocal is required to maintain consistent timbre and personal style (e.g., pronunciation, articulation skills). 
The Flow-based Style Predictor is used to predict the target style $z_s$, learning both personalized styles from $\tilde{z_s}$ and style control information from $z_p$ (such as singing methods).
Notably, $\tilde{z_s}$ and $z_p$ can be input individually for full style control.

For music style transfer, AccompBand uses the noisy prompt accompaniment $\tilde{y_a}$ with a time step 0.5 instead of $\epsilon$ and sums it with target vocal $y_v$, enabling the model to learn the style from the retained components of the prompt accompaniment.

In the vocal-to-song task, the GT vocal is used to guide AccompBand in generating the accompaniment. 
In contrast, for the accompaniment-to-song task, notes $n$ are extracted from the GT accompaniment $\hat{y_a}$ using ROSVOT \citep{li2024robust} to guide VocalBand in vocal generation, while AccompBand is not used.

\section{VocalBand Details}
\label{app: vocal}

\subsection{Model Configuration}
\label{app: vocal_config}

We list the architecture and hyperparameters of VocalBand in Table \ref{tab: vocal}.

\begin{table}[t]
\centering
\small
\begin{tabular*}{\hsize}{l|c|c}
\toprule
\multicolumn{2}{c|}{\bfseries{Hyper-parameter}}                         & \bfseries{Value}    \\
\midrule
\multirow{5}{*}{\shortstack{Phoneme\\ Encoder}}          & Phoneme Embedding         & 256   \\
                                            & Encoder Layers            & 4     \\
                                            & Encoder Hidden            & 256   \\
                                            & Encoder Conv1D Kernel     & 9     \\
                                            & Encoder Conv1D Filter Size& 1024  \\
\midrule
\multirow{2}{*}{\shortstack{Note \\Encoder}} & Pitch Embedding         & 256   \\
                                            & Duration Hidden           & 256   \\
\midrule
\multirow{3}{*}{\shortstack{Timbre \\Encoder}} & Encoder Layers       & 5   \\
                                            & Hidden Size           & 256   \\
                                            & Conv1D Kernel           & 31   \\
\midrule
\multirow{3}{*}{\shortstack{Residual \\Style \\Encoder}}   & Conv Layers       & 5     \\
                                            & RQ Codebook Size          & 256   \\
                                            & Depth of RQ               & 4     \\
\midrule
 \multirow{5}{*}{\shortstack{Flow-based \\Style \\Predictor}} & Conv Layers     & 20     \\
                                             & Kernel Size               & 3      \\
                                             & Residual Channel          & 256    \\
                                             & Hidden Channel            & 256    \\
                                             & Training Time Steps            & 100    \\
\midrule
 \multirow{5}{*}{\shortstack{Flow-based \\Pitch \\Predictor}} & Conv Layers     & 12     \\
                                             & Kernel Size               & 3      \\
                                             & Residual Channel          & 192    \\
                                             & Hidden Channel            & 256    \\
                                             & Training Time Steps            & 100    \\
\midrule
\multirow{5}{*}{\shortstack{Flow-based \\Mel \\Decoder}}    & Conv Layers  & 20     \\
                                             & Kernel Size               & 3      \\
                                             & Residual Channel          & 256    \\
                                             & Hidden Channel            & 256    \\
                                             & Training Time Steps            & 100      \\
\midrule
\multicolumn{2}{c|}{Total Number of Parameters} & 56.26M \\
\bottomrule
\end{tabular*}
\caption{Hyper-parameters of VocalBand.}
\label{tab: vocal}
\end{table}

\subsection{Decomposition Strategy}
\label{app: decom}

We assume that the target vocal $y_v$ can be decomposed into three distinct representations: content $z_c$, style $z_s$ (e.g., singing methods, emotion, techniques, pronunciation, and articulation skills), and timbre $z_t$. 
When a vocal prompt $\tilde{y_v}$ is provided during training, our goal is to transfer both the timbre $\tilde{z_t}$ and personalized style $\tilde{z_s}$ (like pronunciation and articulation skills) from the vocal prompt to the target vocal $y_v$.
Meanwhile, we also need to achieve style control from text tokens $z_p$ (such as singing method, emotion, and techniques).

Following previous style transfer approaches \citep{jiang2024mega}, we assume that the mutual information between $y_v$ and $\tilde{y_v}$ primarily captures global information, represented by $z_t$ (timbre). 
Therefore, the target timbre $z_t$ is set equal to the prompt timbre $\tilde{z_t}$, as we aim to control the timbre of the output based on the user's input.
Under this assumption, $\tilde{z_t}$ is extracted using a timbre encoder, which focuses solely on timbre information, without capturing style $z_s$ or content $z_c$. 
To ensure that the content encoders extract only content-related information, we feed it phoneme sequences and musical notes, allowing it to exclusively pass the content representation $z_c$.
For the timbre and content encoders, please refer to Appendix \ref{app: timbre} and \ref{app: content}.

Once both $z_c$ and $z_t$ are obtained, we must remove fine-grained content and timbre information from the target style $z_s$. 
We employ a residual style encoder to extract the prompt style $\tilde{z_s}$, and then use the Flow-based Style Predictor to predict the target style $z_s$.
The latent vector $z_s$ generated by the Flow-based Style Predictor not only captures the personalized styles consistent with the prompt style $\tilde{z_s}$ (e.g., pronunciation and articulation skills) but also incorporates the styles in the text tokens $z_p$ (like singing methods, emotions, and techniques).

By utilizing a residual quantization (RQ) model \citep{lee2022autoregressive} as an information bottleneck \citep{qian2019autovc}, the residual style encoder is compelled to transmit only the fine-grained style information $z_s$ \citep{zhang2024stylesinger}, which other encoders cannot capture. 
Both $z_s$ and $\tilde{z_s}$ share the same form as the RQ embeddings, consisting of multiple layers of fine-grained style information that are disentangled from both timbre and content. 
This is because $z_s$ is the output of the flow-matching ODE solver, whose training objective is to capture the target style from the ground truth vocals, as extracted by the residual style encoder. 
For more details about the Flow-based Style Predictor, please refer to Appendix \ref{app: fsp}.
Consequently, the process guarantees the successful decomposition of style from content and timbre.
These embeddings  $z_c$, $z_t$, and $z_s$ are then fed into a duration predictor \citep{ren2020fastspeech} and a length regulator for subsequent F0 and mel-spectrogram prediction.

\subsection{Timbre Encoder}
\label{app: timbre}

The timbre encoder, designed to capture the unique identity of the singer, extracts a global timbre vector $\tilde{z_t}$ from the vocal prompt $\tilde{y_v}$. 
The encoder consists of multiple stacked convolutional layers. 
To ensure stability in the timbre representation, the output of the timbre encoder is temporally averaged, producing a one-dimensional timbre vector $\tilde{z_t}$.
The target timbre $z_t$ is set equal to the prompt timbre $\tilde{z_t}$, as we aim to control the timbre of the output based on the user's input.

\subsection{Content Encoders}
\label{app: content}

Our content encoders consist of a phoneme encoder and a note encoder. 
The phoneme encoder processes a sequence of phonemes $p$ through a phoneme embedding layer followed by four FFT blocks, extracting phoneme features. 
In parallel, the note encoder handles musical score information $n$, processing note pitches and durations. 
These are passed through two separate embedding layers and a linear projection layer, which generate the corresponding note features.
The outputs of the phoneme encoder and the note encoder are then summed as $z_c$.

\begin{figure*}[t]
\centering
\includegraphics[width=1\textwidth]{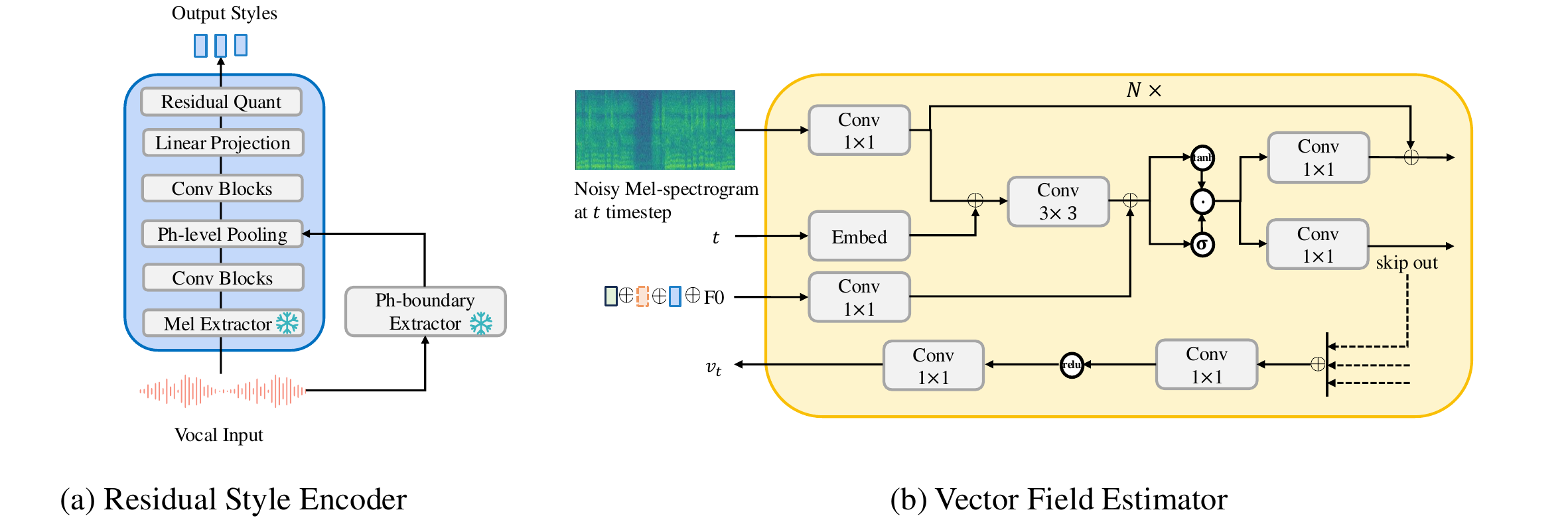}
\caption{The architecture of two components of VocalBand, Figure (a) shows the residual style encoder while Figure (b) illustrates the vector field estimator of the Flow-based Mel Decoder.
}
\label{fig: arch_app}
\end{figure*}

\subsection{Residual Style Encoder}
\label{app: style}

Singing style can vary across and within phonemes. 
To comprehensively capture phoneme-level styles (such as singing methods, emotion, techniques, pronunciation, and articulation skills) and disentangle them from timbre and content, we design the residual style encoder. 
In the residual style encoder, we employ a Residual Quantization (RQ) module \citep{lee2022autoregressive} to extract singing style, creating an information bottleneck that effectively filters out non-style information \citep{zhang2024stylesinger}. 
Thanks to the RQ's ability to extract multiple layers of information, it enables more comprehensive modeling of style across various hierarchical levels. 
Specifically, pronunciation and articulation skills encompass pitch transitions between musical notes and vibrato within a phoneme, where the multi-level modeling capability of RQ is highly suitable.

More concretely, as illustrated in Figure \ref{fig: arch_app} (a), we first extract the mel-spectrogram from the input vocal using the open-source tool librosa \footnote{\label{foot: librosa}\url{https://github.com/librosa/librosa}} and further refine it through convolutional blocks. 
The output is then condensed into phoneme-level hidden states via a pooling layer, which operates based on phoneme boundaries. 
We utilize open-source tools including WhisperX \citep{bain2022whisperx} and Montreal Forced Aligner (MFA) \citep{mcauliffe2017montreal} to extract these phoneme boundaries directly from the input vocal.
Subsequently, the convolution stacks capture phoneme-level correlations.
Next, we use a linear projection to map the output into a low-dimensional latent variable space for code index lookup, significantly enhancing the utilization of the codebook \citep{yu2021vector}. 

With a quantization depth of $n$, the RQ module represents the input $z_e$ as a sequence of $N$ ordered codes. 
Let $RQ_i(z_e)$ denote the process of representing $z_e$ as RQ code and extracting the code embedding in the $i$-th codebook. 
The representation of $z_e$ in the RQ module at depth $n \in [N]$ is denoted as $\hat{z_e}^n = \sum_{i=1}^n RQ_i(z_e)$.
To ensure that the input representation adheres to a discrete embedding, a commitment loss \citep{lee2022autoregressive} is employed:
\begin{equation}
\label{equ: commit}
\begin{aligned}
&\mathcal{L}_{commit} = \sum_{n=1}^N \left\| z_e - sg[\hat{z_e}^n] \right\|_{2}^{2},
\end{aligned}
\end{equation}
where the notation $sg$ represents the stop-gradient operator. 
It is important to note that $\mathcal{L}_{commit}$ is the cumulative sum of quantization errors across all $n$ iterations, rather than a single term. 
The objective is to ensure that $\hat{z_e}^n$ progressively reduces the quantization error of $z_e$ as the value of $n$ increases.
Finally, we extract the phoneme-level style embedding from the input vocal.

\subsection{Flow-based Style Predictor}
\label{app: fsp}

As shown in Figure \ref{fig: arch2} (a), the Flow-based Style Predictor uses content $z_c$, timbre $z_t$, phoneme-level prompt style $\tilde{z_s}$, and text tokens $z_p$ to predict the target style $z_s$. 
With the combined $z_c$ and $z_t$, we employ a style alignment model utilizing the Scaled Dot-Product Attention mechanism \citep{vaswani2017attention} to align style control information from $z_p$ (e.g., singing methods, emotions, techniques) with the content. 
Positional embedding is applied before feeding $z_p$ into the attention module. 
In the attention module, the combined $z_c$ and $z_t$ serve as the query $z_{ct}$, while $z_p$ serves as both the key and value, and $d$ represents the dimensionality of the key and query:
\begin{equation}
\begin{aligned}
&Attention(Q, K, V) = Attention(z_{ct}, z_p, z_p) \\
& = Softmax\left(\frac{z_{ct} z_p^{T}}{\sqrt{d}}\right) z_p.
\end{aligned}
\end{equation}
We stack the style alignment layer multiple times for better performance and gradually stylize the query value.
We combine the output with $z_{ct}$ as condition $c$ and then feed it into an ODE solver, which transforms Gaussian noise $\epsilon$ into $z_s$ along a probability path $p_t(z_{st})$. 
We concatenate $\tilde{z_s}$ with $\epsilon$ to allow $z_s$ to learn personalized styles (e.g., pronunciation and articulation skills). 

During training, we set $u(z_{st}, t)$ to represent the target vector field at time $t$, obtained through linear interpolation between $\epsilon$ and the ground truth (GT) phoneme-level style $z_s$, which is extracted from the GT vocal by the residual style encoder.
To stabilize the flow-matching training process, we do not train the Flow-based Style Predictor during the early stages of training (the first 50,000 steps). 
Instead, we feed the GT style $z_s$ into the subsequent Flow-based Pitch Predictor and Mel Decoder. 
Therefore, by the time we begin training the Flow-based Style Predictor, the residual style encoder has stabilized, ensuring a consistent GT $z_s$, which is beneficial for the flow-matching training.

The learned vector field $v(z_{st}, t | c; \theta)$, predicted by a vector field estimator at each time $t$, ensures smooth interpolation between the initial noise and the output $z_s$, guided by the flow-matching objective.
We use the non-causal WaveNet architecture \citep{van2016wavenet} as the backbone of our vector field estimator, due to its proven capability in modeling sequential data.
For more details about the vector field estimator, please refer to Appendix \ref{app: vec}.
Notably, to enable the model to handle cases without a vocal prompt, we drop vocal prompts with a probability of 0.2 during training. 
We also replace $z_p$ with embedded empty strings in a probability of 0.1 for cases without prompts.

During inference, the ODE solver generates the phoneme-level target style $z_s$ directly from the concatenation of Gaussian noise and $\tilde{z_s}$ (if a vocal prompt is provided), based on the condition $c$.
This method ensures fast and controllable generation of $z_s$, learning personalized styles consistent with $\tilde{z_s}$ while incorporating the aligned style control information from $z_p$.

\subsection{Vector Field Estimator}
\label{app: vec}

We adopt the non-causal WaveNet architecture \citep{van2016wavenet} as the backbone of our vector field estimators for the Flow-based Style Predictor, Pitch Predictor, and Mel Decoder, due to its demonstrated effectiveness in modeling sequential data. 
The architecture of the vector field estimator for the Flow-based Mel Decoder is depicted in Figure \ref{fig: arch_app} (b). 
We input content $z_c$, timbre $z_t$, style $z_s$, and F0 as conditioning factors to predict the corresponding vector field.
Similarly, the architecture of the vector field estimators for the Flow-based Pitch Predictor and Style Predictor follows the same structure, while the only difference lies in the input and condition for each model.

\begin{figure}[t]
  \centering
    \includegraphics[scale=0.4]{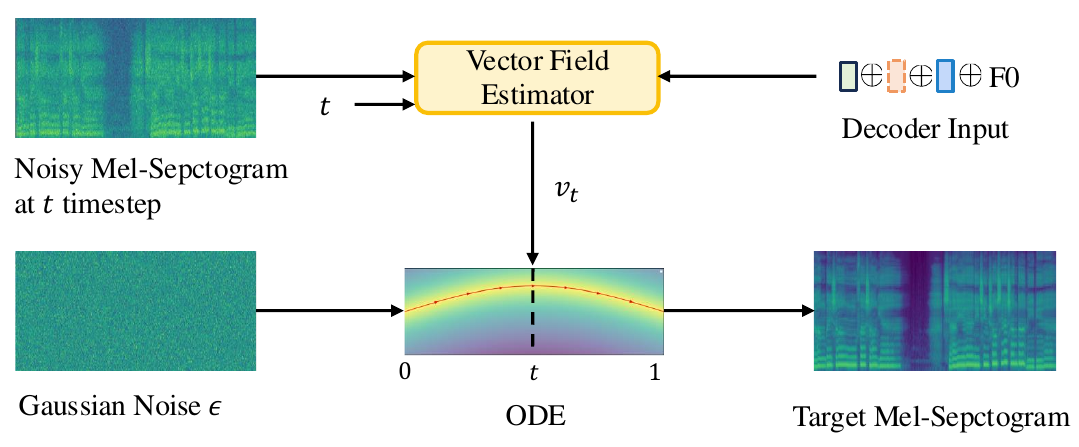}
  \caption{The architecture of Flow-based Mel Decoder.}
  \label{fig: arch_app3}
\end{figure}

\subsection{Flow-based Pitch Predictor and Mel Decoder}
\label{app: decode}

During training, our target F0 is extracted using the open-source tool RMVPE \citep{wei2023rmvpe}, while mel-spectrograms are extracted using the open-source tool librosa \textsuperscript{\ref{foot: librosa}}. 
As shown in Figure \ref{fig: arch_app3}, the Flow-based Mel Decoder employs a flow-matching architecture \citep{liu2022flow}, where the vector field estimator and ODE solver generate the target mel-spectrogram from Gaussian noise $\epsilon$. 
The Flow-based Pitch Predictor follows a similar flow-matching procedure.
We adopt the non-causal WaveNet architecture \citep{van2016wavenet} as the backbone of our vector field estimator. 
For further details on the vector field estimator, please refer to Appendix \ref{app: vec}.

\section{AccompBand Details}
\label{app: accomp}

\subsection{Model Configuration}
\label{app: accomp_config}

We list the architecture and hyperparameters of AccompBand in Table \ref{tab: accomp}.

\begin{table}[t]
\centering
\small
\begin{tabular*}{\hsize}{l|c|c}
\toprule
\multicolumn{2}{c|}{\bfseries{Hyperparameter}}                 & \bfseries{Value}    \\
\midrule
 \multirow{4}{*}{\shortstack{Accomp \\Encoder}}    & Encoder Layers            & 3     \\
                                            & Encoder Hidden            & 384   \\
                                            & Encoder Conv1D Kernel     & 5     \\
                                            & Encoder Output Channels   & 20     \\
\midrule
 \multirow{4}{*}{\shortstack{Accomp \\Decoder}}   & Decoder Layers            & 3     \\
                                            & Decoder Hidden            & 384   \\
                                            & Decoder Conv1D Kernel     & 5     \\
                                            & Decoder Input Channels   & 20     \\
\midrule
 \multirow{4}{*}{\shortstack{Vocal \\Encoder}}   & Encoder Layers            & 3     \\
                                            & Encoder Hidden            & 384   \\
                                            & Encoder Conv1D Kernel     & 5     \\
                                            & Encoder Output Channels   & 20     \\
\midrule
 \multirow{5}{*}{\shortstack{Band \\Transformer \\Blocks}} & Transformer Layers & 4  \\
                                            & Transformer Embed Dim     & 768     \\
                                            & Transformer Attention Headers  & 8   \\
                                            & Experts for each group    & 4   \\
                                            & Training Time Steps            & 1000    \\
\midrule
\multicolumn{2}{c|}{Total Number of Parameters} & 431.07M \\
\bottomrule
\end{tabular*}
\caption{Hyper-parameters of AccompBand.}
\label{tab: accomp}
\end{table}

\subsection{Accomp Encoder and Decoder}
\label{app: align}


The accomp encoder and decoder are based on the VAE model \citep{kingma2013auto}. 
For pre-training the accomp encoder and decoder, we use the L2 reconstruction loss: $\mathcal{L}_{rec}=\|y_{v}-\hat{y_v}\|^2,$where $y_v$ is the reconstructed accompaniment mel-spectrogram and $\hat{y_v}$ is the ground truth accompaniment mel-spectrogram. 
Additionally, we incorporate a GAN discriminator, following the architecture of ML-GAN \citep{chen2020hifisinger}, to further enhance the quality of the reconstruction. 
We apply the LSGAN-style adversarial loss \citep{mao2017least}, $\mathcal{L}_{adv}$, which aims to minimize the distributional distance between the predicted mel-spectrograms and the ground truth mel-spectrograms.
Before feeding the waveform into the accomp encoder, we first extract the mel-spectrogram using librosa \textsuperscript{\ref{foot: librosa}}. 
After generating the mel-spectrogram from the decoder output, we utilize HiFi-GAN \citep{kong2020hifi} to convert it back into audio.

\subsection{Band Transformer Blocks}
\label{app: trans}
As shown in Figure \ref{fig: arch} (c), the Band Transformer Blocks are based on Flag-Dit \citep{gao2024lumina}. 
During training, the vocal embedding $z_v$ extracted by the vocal encoder is added to the noisy input $x_t$ to leverage the transformer's self-attention mechanism, allowing the model to learn vocal-matching style, rhythm, and melody. 
We use RMSNorm \citep{zhang2019root} to improve training stability, preventing the absolute values from growing uncontrollably and causing numerical instability.
Next, we compute the global style embedding $z_g$ by averaging the text tokens $z_p$ and vocal embedding $z_v$ along the temporal dimension and adding the time step embedding of $t$. 
This global style embedding is used in a multi-layer style adaptor, which modulates the latent representation using adaptive layer normalization (AdaLN) \citep{peebles2023scalable} to ensure style consistency. 
We compute the scale and shift using linear regression:
\begin{equation}
\begin{aligned}
&AdaLN(h, c) = \gamma_c \times LayerNorm(h) + \beta_c,
\end{aligned}
\end{equation}
where $h$ represents the hidden representation. 
We zero-initialize the batch norm scale factor $\gamma$ in each block \citep{peebles2023scalable}. 
Moreover, we explore relative positional encoding with rotary positional embedding (RoPE) \citep{su2024roformer}, which injects temporal positional information into the model. 
This enables the model to capture the temporal relationships between successive frames, providing significant performance improvements for the transformer.

Then, the zero-initialized attention mechanism \citep{bachlechner2021rezero} is used to inject conditional information from the text tokens $z_p$ into the hidden states $h$, while simultaneously learning the vocal style, rhythm, and melody aligned with the vocal embedding $z_v$ added to $x_t$. 
Given the accompaniment queries $Q_h$, keys $K_h$, and values $V_h$ from hidden states, along with the text keys $K_z$ and values $V_z$, the final attention output is:
\begin{equation}
\begin{aligned}
&Attention = softmax \left( \frac{\tilde{Q_h} \tilde{K_h}^\top}{\sqrt{d}} \right) V_h +\\
&\tanh(\alpha) softmax \left( \frac{\tilde{Q_h} K_z^\top}{\sqrt{d}} \right) V_z,
\end{aligned}
\end{equation}
where $\tilde{Q_h}$ and $\tilde{K_h}$ denote using RoPE in both queries and keys, $d$ is the dimensionality of both queries and keys, and $\alpha$ is a zero-initialized learnable parameter that gates the cross-attention with the input text tokens. 

\subsection{Band-MOE}
\label{app: moe}

As illustrated in Figure \ref{fig: arch2}(b), Band-MOE is composed of three expert groups: Aligned MOE, Controlled MOE, and Acoustic MOE, each comprising multiple experts. 
We employ Feed-Forward Networks (FFNs) as the architecture for each expert.
It is well-established \citep{lee2022bigvgan} that mel-spectrogram details exhibit different patterns across various acoustic frequencies. 
In musical accompaniment, high-frequency components often include the harmonics and overtones of instruments like strings and flutes, as well as percussive elements such as cymbals and hi-hats, which enhance the brightness and clarity of the sound. 
Conversely, low-frequency content encompasses basslines and kick drums, providing foundational rhythm and depth that shape the overall groove and warmth of the music.
Motivated by this, previous works \citep{kong2020hifi} have adopted multi-scale architectures to model downsampled signals at different frequency bands, which effectively control the periodic elements of the signal and reduce artifacts.

Building on this idea, we introduce Acoustic MOE, where experts are assigned to specific acoustic frequency bands based on the processed hidden representation $h$, and their outputs are aggregated to produce the final result. 
Moreover, since the vocal and accomp encoder employ 1D convolutions to encode both the vocal and accompaniment mel-spectrograms, the latent representation of the hidden $h$ should retain the frequency distribution.

Our routing strategy for all routers is based on the dense-to-sparse Gumbel-Softmax method \citep{nie2021evomoe}, enabling dynamic and efficient expert selection.
The Gumbel-Softmax trick facilitates sampling from a categorical distribution by reparameterizing categorical variables to make them differentiable. 
Specifically, the routing score $g(h)$ for each expert $i$ is computed as follows:
\begin{equation}
\begin{aligned}
&g(h)_i = \frac{\exp((h \cdot W_g + \zeta_i) / \tau)}{\sum_{j=1}^N \exp((h \cdot W_g + \zeta_j) / \tau)},
\end{aligned}
\end{equation}
where $W_g$ is the learned gating weight, $\zeta$ is sampled from the Gumbel(0, 1) distribution \citep{jang2016categorical}, and $\tau$ is the softmax temperature. 
Initially, a high temperature $\tau$ results in denser expert selection, allowing multiple experts to process the same input. 
As training progresses, $\tau$ is gradually decreased, making the routing sparser and selecting fewer experts for each input. 
When $\tau \to 0$, the distribution approaches a nearly one-hot form, effectively selecting the most suitable expert for each token.
Following prior work \citep{nie2021evomoe}, we dynamically reduce $\tau$ from 2.0 to 0.3 during training and use the hard mode during inference, selecting only one expert.
Notably, only the global router does not conduct hard mode during inference, as we need experts from different expert groups to cooperate in accompaniment generation.
The algorithm of Band-MOE is shown in Algorithm \ref{alg: moe}.

\renewcommand{\algorithmicrequire}{\textbf{Input:}}
\renewcommand{\algorithmicensure}{\textbf{Output:}}
\begin{algorithm*}[t]
\caption{Pseudo-Code of Band-MOE Routing Strategy}
\label{alg: moe}
\begin{algorithmic}[1]
\REQUIRE Input hidden representation $h$, vocal embedding $z_v$, text prompt embedding $z_p$, time step $t$
\ENSURE Output with enhanced quality and control $o_{\text{final}}$
\STATE Initialize Gumbel-Softmax temperature $\tau$, sample Gumbel noise $\zeta$
\FOR{each time step $t$}
    \STATE \textbf{Aligned MOE:}
    \STATE \quad Use Gumbel-Softmax for each token in the time channel to select an expert by $z_v$:
    \STATE \quad $g_{\text{aligned}}(h) \gets \text{GumbelSoftmax}(z_v \cdot W_{\text{aligned}} + \zeta) / \tau$
    \STATE \quad Compute Aligned MOE output:
    \STATE \quad $o_{\text{aligned}} \gets \sum_{i} g_{\text{aligned}, i} \cdot \text{Expert}_{i, \text{aligned}}(h)$
    \STATE \textbf{Controlled MOE:}
    \STATE \quad Use Cross-Attention extracting style for alignment between $z_p$ and $h$:
    \STATE \quad $z_{sty} \gets \text{CrossAttention}(h(Q),z_p(K),z_p(V))$
    \STATE \quad Use Gumbel-Softmax for each token in the time channel to select an expert by $z_{sty}$:
    \STATE \quad $g_{\text{controlled}}(h) \gets \text{GumbelSoftmax}(z_{sty} \cdot W_{\text{controlled}} + \zeta) / \tau$
    \STATE \quad Compute Controlled MOE output:
    \STATE \quad $o_{\text{controlled}} \gets \sum_{i} g_{\text{controlled}, i} \cdot \text{Expert}_{i, \text{controlled}}(h)$
    \STATE \textbf{Global Router:}
    \STATE \quad Use Gumbel-Softmax to compute global weights $\alpha_t$ and $\beta_t$:
    \STATE \quad $g_{\text{global}}(h) \gets \text{GumbelSoftmax}(embedding(t) \cdot W_{\text{global}} + \zeta) / \tau$
    \STATE \quad $\alpha_t, \beta_t \gets g_{\text{global}}(h)$
    \STATE \quad Combine Aligned and Controlled MOE outputs:
    \STATE \quad $o_{\text{combined}} \gets \alpha_t \cdot o_{\text{aligned}} + \beta_t \cdot o_{\text{controlled}}$
    \STATE \textbf{Acoustic MOE:}
    \STATE \quad Use Gumbel-Softmax to select an expert for each frequency channel:
    \STATE \quad $g_{\text{acoustic}}(o_{\text{combined}}) \gets \text{GumbelSoftmax}(o_{\text{combined}} \cdot W_{\text{acousitc}} + \zeta) / \tau$
    \STATE \quad Compute Acoustic MOE output:
    \STATE \quad $o_{\text{acoustic}} \gets \sum_{j} g_{\text{acoustic}, j} \cdot \text{Expert}_{j, \text{acoustic}}(o_{\text{combined}})$
\ENDFOR
\STATE Return $o_{\text{final}} \gets o_{\text{acoustic}}$ as the final routed output
\end{algorithmic}
\end{algorithm*}

Moreover, to avoid overloading any individual expert and ensure balanced utilization, we incorporate a load-balancing loss for each expert group \citep{fedus2022switch}. 
The balance loss $ \mathcal{L}_{balance} $ is:
\begin{equation}
\label{equ: balance}
\begin{aligned}
&\mathcal{L}_{balance} = \alpha N \sum_{i=1}^{N} \left( \frac{1 }{B} \sum_{h \in B} g(h)_i \right).
\end{aligned}
\end{equation}
where $B$ is the batch size, $N$ is the number of experts, and $\alpha$ is a hyperparameter controlling the strength of the regularization, for which we use 0.1.
This loss encourages a more uniform distribution of tokens across experts, improving training efficiency by preventing expert underutilization or overload.
Thus, our routing strategy not only allows dynamic expert selection but also ensures that the computational load is evenly distributed across experts, reducing training time and improving the model performance of Band-MOE. 

\subsection{Classifier-free Guidance}
\label{app: cfg}

During AccompBand training, we randomly replace the input text tokens with embedded empty strings at a probability of 0.2. 
The empty strings are processed through the text encoder to extract text tokens and are padded to a fixed length, like the original text prompts, like previous works \citep{jiang2025sparse}.
For $\gamma$ in Equation \ref{equ: cfg}, a higher $\gamma$ emphasizes the control of the text prompt, improving generation quality by making the outputs more aligned with the given conditions. 
In contrast, a lower $\gamma$ allows for more diverse outputs by reducing the reliance on the text prompt, though this may result in lower relevance to the input prompt. 
In our major accompaniment generation experiments, we use $\gamma = 3$.

\section{LyricBand and MelodyBand}
\label{app: other}

\subsection{LyricBand}
\label{app: lyric}

To enhance the customizability of our song generation system, we introduce LyricBand, a model designed to generate complete song lyrics based on arbitrary text prompts. 
Users can input parameters such as theme, emotion, genre, style, and specific keywords to generate fully personalized lyrics tailored to their preferences.
To effectively train LyricBand, we leverage GPT-4o \citep{achiam2023gpt} to extract prompts from a large corpus of existing song lyrics in our training data. 
These prompts encapsulate essential elements such as the thematic content, emotional tone, narrative perspective, rhyme scheme, and stylistic features of the songs. 
By extracting this rich set of attributes, we create a comprehensive dataset that pairs textual prompts with corresponding lyrics, enabling the model to learn the mapping between user inputs and desired lyrical outputs.

We employ QLoRA \citep{dettmers2024qlora} for efficient fine-tuning of the well-performing open-source bilingual large language model Qwen-7B \citep{bai2023qwen}. 
By utilizing 4-bit quantization and low-rank adapters, QLoRA significantly reduces the computational resources required for fine-tuning while preserving the model's performance. 
This approach allows LyricBand to adapt effectively to the task of lyrics generation, maintaining high levels of customization and creativity across a diverse range of user prompts.
In our experiments, we set LoRA $r=32,\alpha=16$.
LyricBand demonstrates the capability to capture nuanced themes and emotions specified by users, generating lyrics that not only align with the given prompts but also exhibit coherent structure and artistic expression.

\begin{table*}[t]
\small
\centering
\begin{tabular}{l|cccccc}
\toprule
\bfseries{Dataset}                             & Type & Languages  & Annotation    & Duration (hours)  \\
\midrule
GTSinger \citep{zhang2024gtsinger}    & vocal  & Chinese \& English & lyrics, notes,styles & 29.6    \\
M4Singer \citep{zhang2022m4singer}    & vocal  & Chinese & lyrics, notes & 29.8    \\
OpenSinger \citep{huang2021multi}     & vocal  & Chinese& lyrics & 83.5      \\
LP-MusicCaps-MSD \citep{doh2023lp}    & accomp  & / & text prompt        & 213.6       \\
web-crawled                         & song & Chinese, English & /  & 979.4 \\
\bottomrule 
\end{tabular}
\caption{Statistics of training datasets.}
\label{tab: data}
\end{table*}

\subsection{MelodyBand}
\label{app: melody}

Previous singing voice and song generation models often require users to provide music scores to achieve stable melodies \citep{zhiqing2024text}, lacking personalized customization of the melody. 
Inspired by symbolic music generation models \citep{dong2018musegan}, we introduce MelodyBand, an additional model where melody-related features like notes are generated from text descriptions in advance.
By using notes as the representation of the melody, we can achieve more stable melody control. 
However, requiring users to provide music scores is impractical. 
Generating notes using natural language prompts can both ensure stable melodies and allow for flexible customization. 
For controllable melody generation, we construct artificial textual prompts to deliver melody-related information. 
Musical attributes like key, tempo, vocal range, and other information can be used as prompts for melody customization.

\begin{figure}[t]
  \centering
    \includegraphics[scale=0.5]{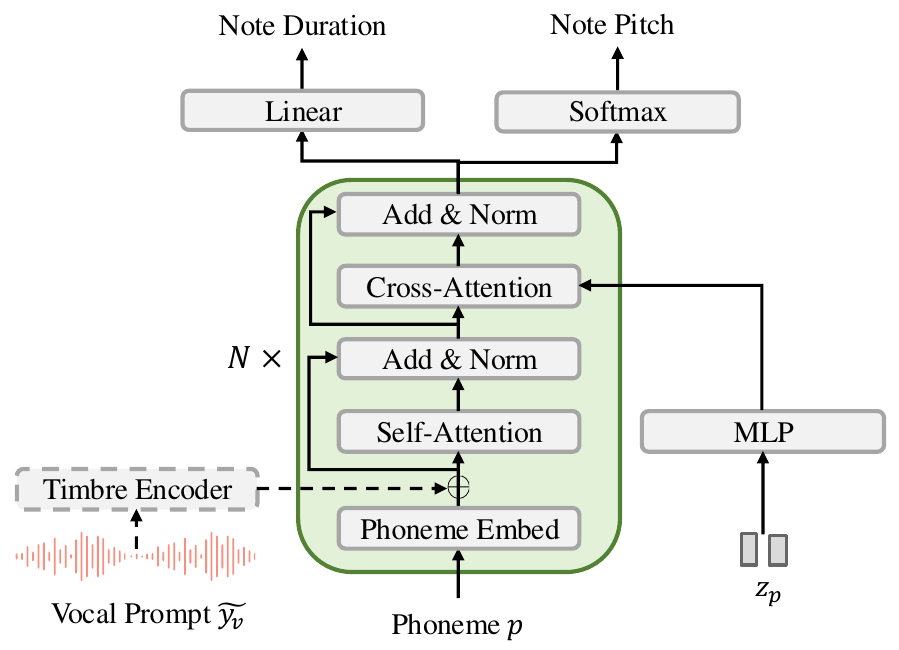}
  \caption{The architecture of MelodyBand.}
  \label{fig: arch_app2}
\end{figure}

When users do not input music scores, as shown in Figure~\ref{fig: arch_app2}(b), MelodyBand takes the phonemes of the lyrics as content information and optional vocal prompts to extract timbre. 
It composes music for the lyrics and selects appropriate frequencies based on the timbre, using text prompts for style control. 
We employ a non-autoregressive transformer model to efficiently generate note pitches and durations simultaneously. 
The non-autoregressive transformer enables fast and high-quality generation, making it suitable for our multi-task song generation system.

With encoded phonemes and timbre, we inject text prompts through cross-attention transformers, allowing the model to integrate linguistic cues more effectively. 
Several heads are added to generate note pitches and durations.
We pass each dimension of the stacked output through a softmax function to generate note pitches and through a linear layer to generate note durations.
We train MelodyBand using cross-entropy loss for note pitches and an L2 loss for note durations. 
Let the true note pitch and duration for $i$-th phoneme be $n_p^{(i)}$ and $n_d^{(i)}$, and the GT note pitch and duration be $\hat{n}_p^{(i)}$ and $\hat{n}_d^{(i)}$, respectively. 
The cross-entropy loss $\mathcal{L}_{pitch}$ is:
\begin{equation} 
\label{equ: pitch}
\begin{aligned}
&\mathcal{L}_{pitch} = -\sum_{i=1}^N \sum_{k=1}^{K} \delta_{\hat{n_p}^{(i)},k} \log(P_k^{(i)}), 
\end{aligned}
\end{equation}
where $N$ is the length of phoneme sequence, $K$ is number of pitch classes, $\delta_{\hat{n_p}^{(i)},k}$ is 1 if $\hat{n_p}^{(i)} = k$ and 0 otherwise. 
$P_k^{(i)}$ is the predicted probability of pitch $k$ at time $i$. 
The L2 loss $\mathcal{L}_{duration}$ is:
\begin{equation} 
\label{equ: dur}
\begin{aligned}
&\mathcal{L}_{duration} = \sum_{i=1}^N \left( n_d^{(i)} - \hat{n}_d^{(i)} \right)^2.
\end{aligned}
\end{equation}
Our MelodyBand employs 8 transformer layers, and 8 attention heads, the hidden size is 768, with 23.32M parameters in total.

\begin{figure*}[t]
\centering
\includegraphics[width=1\textwidth]{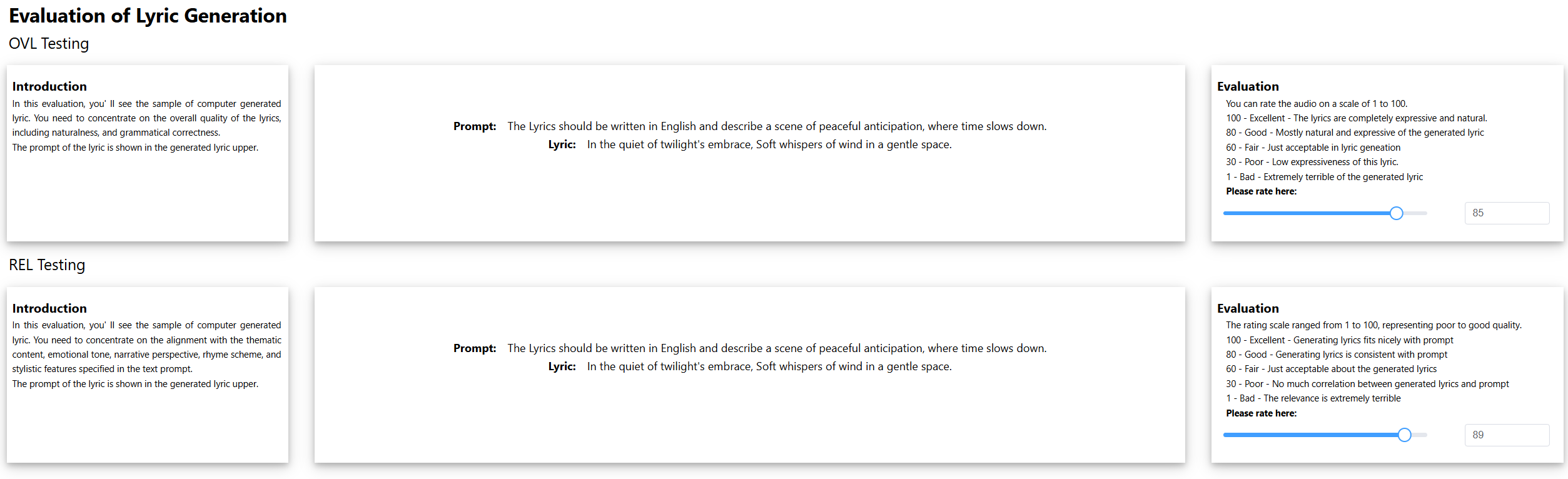}
\caption{Screenshot of lyric evaluation.
}
\label{fig: lyric}
\end{figure*}

\begin{figure*}[t]
\centering
\includegraphics[width=1\textwidth]{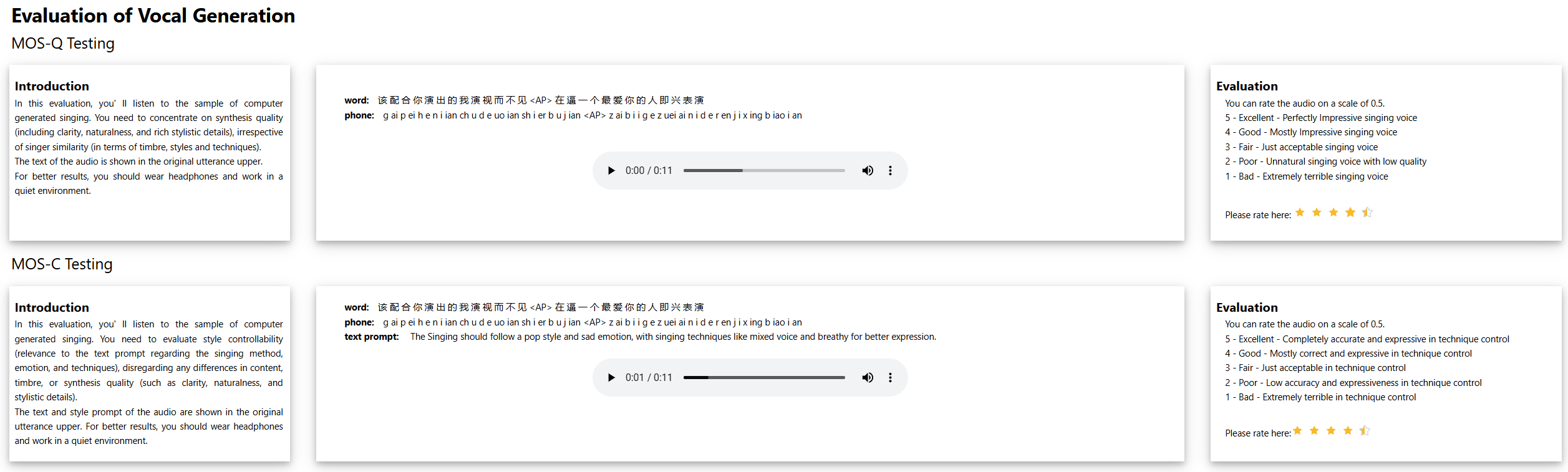}
\caption{Screenshot of vocal evaluation.
}
\label{fig: vocal}
\end{figure*}

\begin{figure*}[t]
\centering
\includegraphics[width=1\textwidth]{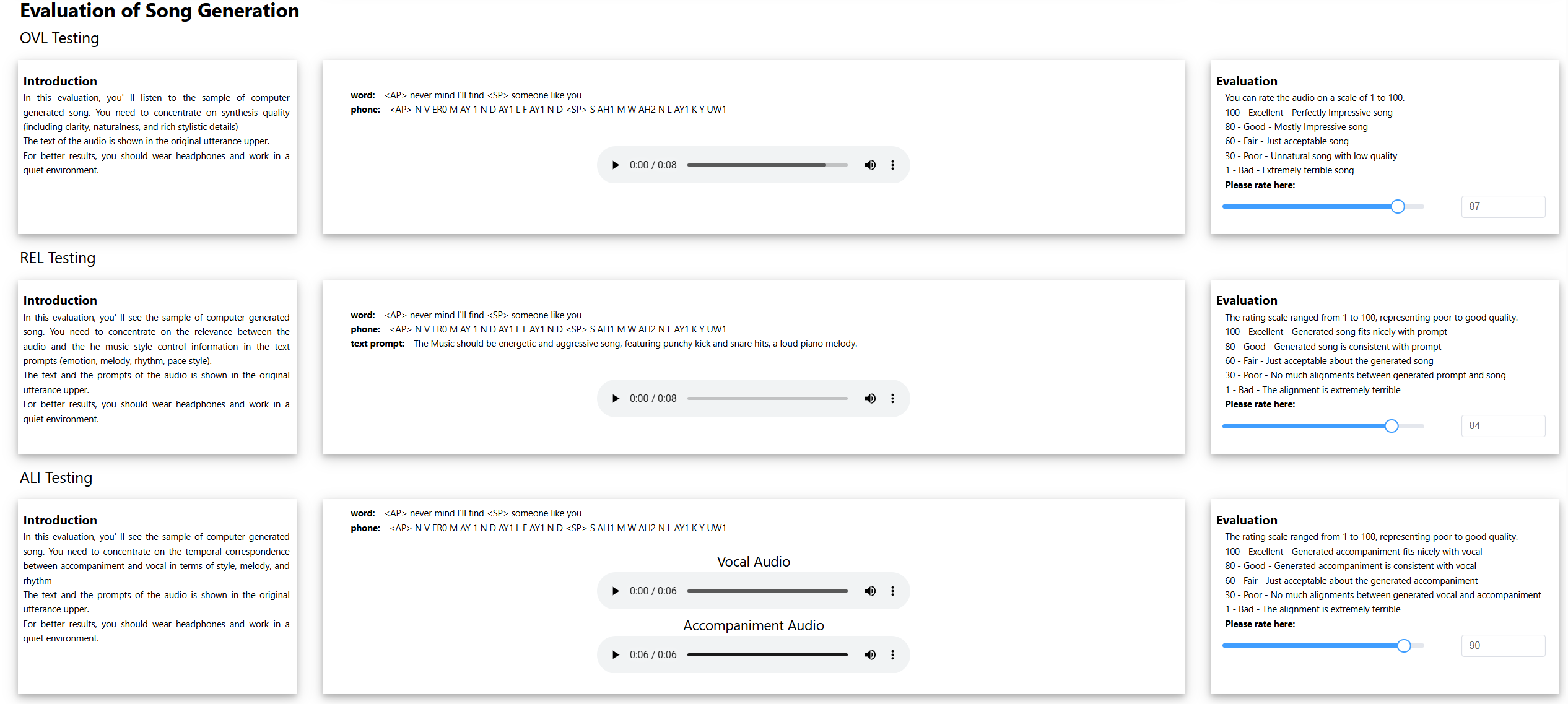}
\caption{Screenshot of song evaluation.
}
\label{fig: song}
\end{figure*}

\section{Dataset Analysis}
\label{app: data}

We train our model using a combination of bilingual web-crawled song datasets and open-source singing datasets. 
Since there are no publicly available annotated song datasets, we collect 20k Chinese and English songs from well-known music websites. 
To expand data, we also use open-source singing datasets including GTSinger \citep{zhang2024gtsinger} (30 hours in Chinese and English), M4Singer \citep{zhang2022m4singer} (30 hours in Chinese), and OpenSinger \citep{huang2021multi} (83 hours in Chinese).
After processing and cleaning, we have about 1,000 hours of song data (about 80\% in Chinese and 20\% in English) and 1,150 hours of vocal data. 
For accompaniment generation, we use a filtered subset of LP-MusicCaps-MSD \citep{doh2023lp}, resulting in a total size of around 1,200 hours.
We use all open-source datasets under license CC BY-NC-SA 4.0.
The statistics of the datasets are listed in \ref{tab: data}. 

For the web-crawled data, we use Ultimate Vocal Remover \footnote{https://github.com/Anjok07/ultimatevocalremovergui}, an open-source music source separation tool, to perform the vocal-accompaniment separation. 
We utilize WhisperX \citep{bain2022whisperx} to automatically transcribe the demixed vocals, and Montreal Forced Aligner (MFA) \citep{mcauliffe2017montreal} is employed for phoneme and vocal alignment. 
After that, we filter the samples using Silero VAD \citep{team2021silero} to eliminate unvoiced clips. 
The samples are segmented into phrases with a maximum length of 20 seconds, resulting in an average segment duration of 12 seconds.

We utilize a music captioning model \citep{doh2023lp} to generate text prompts from the segmented song clips, and GPT-4o \citep{achiam2023gpt} is used to separate music styles (such as genre, tone, and instrumentation) from vocal descriptions (such as emotion and gender).
For singing styles, following style labels of GTSinger, we hire music experts to annotate all songs for the global singing method (e.g., pop or bel canto) and to label around 200 hours of segmented vocal clips for specific techniques used (e.g., mixed voice, falsetto, breathy, vibrato, glissando, and pharyngeal). 
We hire all music experts and annotators with musical backgrounds at a rate of \$300 per hour. 
They have agreed to make their contributions available for research purposes. 
These annotations, along with the separated vocal descriptions, form the complete singing styles.
For melody styles, we extract the key from the segmented demixed vocal clips using music21 \footnote{https://github.com/cuthbertLab/music21}, tempo and duration using librosa \textsuperscript{\ref{foot: librosa}}, and then use GPT-4o to combine these elements, generating natural language descriptions of vocal ranges based on the average pitch.
For lyric styles, we process lyrics using GPT-4o to extract elements such as thematic content, emotional tone, narrative perspective, rhyme scheme, and stylistic features. 

All styles are combined, along with various annotations, to form the final text prompts. 
During generation, we randomly omit certain elements or entire styles to enhance the model's generalization ability.
We utilize ROSVOT \citep{li2024robust} to obtain note sequences from the segmented demixed vocal clips. 
For vocal and accompaniment data that lacks specific annotations, we use corresponding methods to complete the labeling process.

\section{Evaluation Metrics}
\label{app: eva}

\subsection{Lyric and Melody Evaluation}

For evaluating lyric generation, we randomly select 30 prompts and generate 30 sets of lyrics. 
Each set is evaluated by at least 15 raters for overall quality (OVL) and relevance to the prompt (REL) as subjective evaluation metrics. 
The rating scale ranged from 1 to 100, representing poor to good quality. 
OVL focused on the overall quality of the lyrics, including naturalness, and grammatical correctness, while REL assessed the alignment with the thematic content, emotional tone, narrative perspective, rhyme scheme, and stylistic features specified in the text prompt.
All participants are fairly compensated for their time and effort at a rate of \$12 per hour. 
They are also informed that the results will be used for scientific research purposes.
The testing screenshot is shown in Figure \ref{fig: lyric}.

In melody generation, multiple objective metrics are employed to evaluate controllability. 
We use the Krumhansl-Schmuckler algorithm to predict the potential key of the generated notes and report the average key accuracy (KA). 
If the Pearson correlation coefficient of the ground truth (GT) notes corresponding to the GT key is $r$, and the predicted MIDI corresponding to the GT key is $\hat{r}$, we define the key accuracy as $KA = \hat{r}/r$ (only valid if $r \neq 0$).
We also compute the average absolute difference of average pitches (APD) and temporal duration (TD, in seconds). 
Moreover, following previous work \citep{sheng2021songmass}, we record pitch and duration distribution similarity (PD and DD). 
Specifically, we calculate the distribution (frequency histogram) of pitches and durations in notes and measure the distribution similarity between generated notes and ground truth notes:
\begin{equation}
\begin{aligned}
&\frac{1}{N_s}\sum_{i=1}^{N_{{s}}}OA({Dis}_i,\hat{{Dis}}_i),
\end{aligned}
\end{equation}
where ${Dis}_i$ and $\hat{{Dis}}_i$ represent the pitch or duration distribution of the $i$-th generated and ground-truth song, respectively, $N_s$ is the number of songs in the test set, and OA represents the average overlapped area.
Melody distance (MD) is also computed with dynamic time warping (DTW) \citep{berndt1994using}. 
To evaluate the pitch trend of the melody, we spread out the notes into a time series of pitch according to the duration, with a granularity of 1/16 note. 
Each pitch is normalized by subtracting the average pitch of the entire sequence. 
To measure the similarity between generated and ground-truth time series with different lengths, we use DTW to compute their distance.

\subsection{Vocal Evaluation}

For vocal generation, we randomly select 30 pairs of sentences from our test set for subjective evaluation. 
Each pair consists of a ground truth (GT) and a synthesized vocal, each listened to by at least 15 professional listeners. 
For MOS-Q evaluations, these listeners are instructed to focus on synthesis quality (including clarity, naturalness, and richness of stylistic details) without considering the style control relevance to text prompts. 
For MOS-C, the listeners are informed to evaluate style controllability (relevance to the text prompt regarding the singing method, emotion, and techniques), disregarding any differences in content, timbre, or synthesis quality (such as clarity, naturalness, and stylistic details). 
In both MOS-Q and MOS-C evaluations, listeners are asked to grade various vocal samples on a Likert scale from 1 to 5.
For fairness, all samples are resampled to 24kHZ.
The screenshots of instructions are shown in Figure \ref{fig: vocal}.

We employ F0 Frame Error (FFE) to evaluate the test set's synthesis quality objectively.
FFE combines metrics for voicing decision error and F0 error, capturing essential synthesis quality information. 
For comparison with the FFE reported in the MelodyLM paper, we resample all audio to 24kHz.

For singing style transfer, subjective evaluation is conducted using pairs of audio, where each pair includes a prompt vocal and a synthesized vocal. 
During MOS-S evaluations, listeners are asked to assess singer similarity in terms of timbre and personalized styles to the vocal prompt, disregarding any differences in content or synthesis quality. 

To objectively evaluate timbre similarity, we employ Cosine Similarity (Cos). 
Cos measures the resemblance in singer identity between the synthesized vocal and the vocal prompt by computing the average cosine similarity between the embeddings extracted from the synthesized voices and the vocal prompt, thus providing an objective indication of singer similarity performance. 
Specifically, we use a voice encoder \footnote{https://github.com/resemble-ai/Resemblyzer} to extract singer embeddings.

In all MOS-Q, MOS-S, and MOS-C evaluations, listeners are requested to grade the vocal samples on a Likert scale ranging from 1 to 5. 
All participants are fairly compensated for their time and effort. 
We compensate participants at a rate of \$12 per hour. 
Participants are informed that the results will be used for scientific research.

\subsection{Song Evaluation}

For the subjective evaluation of song generation, we randomly select 30 audio samples from our test set. 
Each sample is listened to by at least 15 raters. 
Following previous work \citep{copet2024simple,zhiqing2024text}, we ask human raters to evaluate three aspects of the audio samples: (i) overall quality (OVL), (ii) relevance to the text prompts (REL), and (iii) alignment with the vocal (ALI). 
For the overall quality test, raters are asked to rate the perceptual quality of the provided samples. 
For the text relevance test, raters evaluate how well the audio matches the music style control information in the text prompts. 
For the alignment with the vocal test, raters focus on the temporal correspondence between accompaniment and vocal in terms of style, melody, and rhythm. Ratings are given on a scale from 1 to 100.

All participants are fairly compensated for their time and effort, with a rate of \$12 per hour. 
Participants are informed that the results will be used for scientific research. 
For fairness, all samples are resampled to 24kHZ and normalized to -23dB LUFS.
The screenshots of instructions in the song generation task are shown in Figure \ref{fig: song}.

For the objective evaluation, we use Frechet Audio Distance (FAD), Kullback-Leibler Divergence (KLD), and the CLAP score (CLAP). 
We report the FAD \citep{kilgour2018fr} using the official implementation in TensorFlow with the VGGish model \footnote{https://github.com/google-research/google-research/tree/master/frechet\_audio\_distance}. 
A low FAD score indicates that the generated audio is plausible. 
Following previous work \citep{copet2024simple}, we compute the KL-divergence over the probabilities of the labels between the GT and the generated music. 
Finally, the CLAP score \citep{wu2023large} is computed between the track description and the generated audio to quantify audio-text alignment, using the pre-trained CLAP model \footnote{https://github.com/LAION-AI/CLAP}.

\begin{figure*}[t]
\centering
\includegraphics[width=1\textwidth]{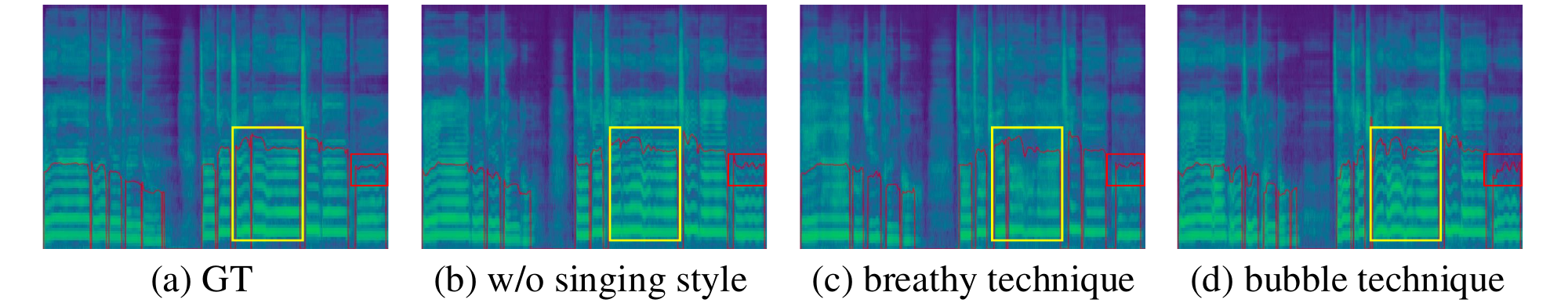}
\caption{Visualization of the mel-spectrogram results generated by VocalBand under different singing styles in the text prompt. The red box contains the fundamental pitch, and the yellow box contains the details of harmonics.
}
\label{fig: svs}
\end{figure*}

\begin{figure*}[t]
\centering
\includegraphics[width=1\textwidth]{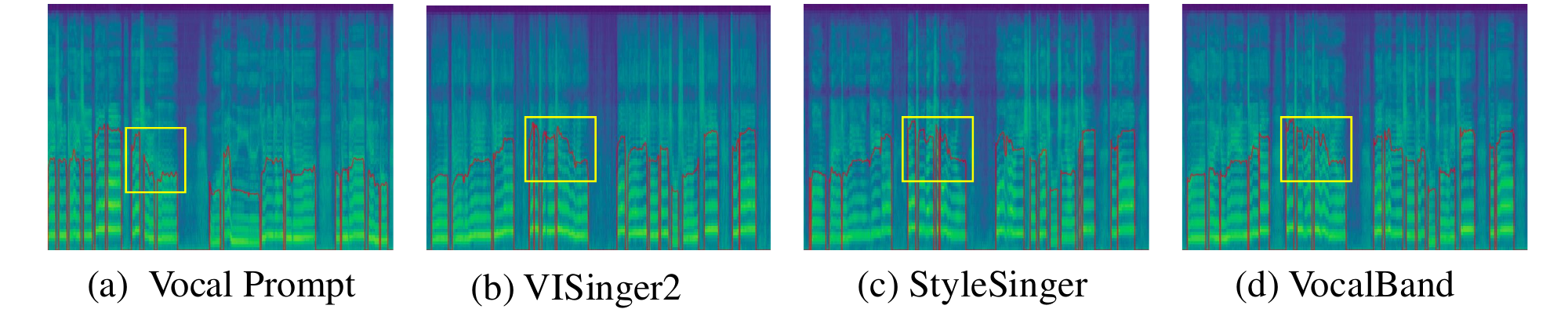}
\caption{Visualization of the mel-spectrogram results generated by VocalBand for singing style transfer. The yellow box contains the fundamental pitch.
}
\label{fig: st}
\end{figure*}

\section{Baseline Models}
\label{app: base}

For lyric generation, we employ the fine-tuned Qwen-7B model as our lyric generation component and use its original (non-fine-tuned) version as the baseline.
For melody generation, we compare against SongMASS \citep{sheng2021songmass} and the MIDI component of MelodyLM \citep{li2024accompanied}, both of which are capable of generating MIDI sequences based on Transformer architectures.

For vocal generation, we compare with VISinger2 \citep{zhang2022visinger}, a traditional high-fidelity SVS model, StyleSinger \citep{zhang2024stylesinger}, the current state-of-the-art zero-shot SVS model with style transfer capabilities. 
We incorporate our text encoder and style alignment modules into the open-source VISinger2 and StyleSinger implementations to enable style control. 
These two models represent well-performing open-source SVS baselines: one follows the traditional SVS paradigm, while the other supports zero-shot style transfer and style modeling. 
Meanwhile, we also compare with vocal parts of Melodist \citep{zhiqing2024text} and MelodyLM.
They both leverage Transformer models for vocal generation, allowing for direct comparisons like other SVS models. 
It should be noted that Melodist lacks zero-shot capabilities and style transfer features. 
Similarly, Melodist and MelodyLM cannot perform natural language prompt–based style control. 

For song generation, we compare with Melodist and MelodyLM, two representative text-to-song models. 
They publicly report datasets of a similar scale to ours and provide comprehensive subjective and objective evaluation metrics for song generation, thus enabling fair comparisons. 
For Melodist and MelodyLM, we rely on their papers and available demos for evaluation. For the other models, we use the corresponding open-source codes.

\section{Multi-Task Experiments}
\label{app: exp}

\begin{table*}[t]
\centering
\small
\begin{tabular}{l|cccc}
\toprule
\textbf{Methods}       & {FAD $\downarrow$} & {KLD $\downarrow$}  & OVL $\uparrow$ &{ALI-A $\uparrow$}\\
\midrule    
VersBand (w/o prompt)  & 3.01 & 1.27 & \bf87.92$\pm$1.73&- \\   
VersBand  & \bf3.02 & \bf1.26 & 87.34$\pm$1.28& \bf80.24$\pm$1.57 \\   
\bottomrule
\end{tabular}
\caption{Results of music style transfer. Prompt means prompt accompaniment.}
\label{tab: mst}                 
\end{table*}

\begin{table*}[t]
\centering
\small
\begin{tabular}{l|cccc}
\toprule
\textbf{Methods}       & {FAD $\downarrow$} & {KLD $\downarrow$}  & OVL $\uparrow$ &{ALI $\uparrow$}\\
\midrule
VersBand (w/o GT)  & 3.01 & 1.27 & 87.92$\pm$1.73&80.51$\pm$1.66 \\
\midrule
MelodyLM  & 3.13 & 1.31  & 84.67$\pm$1.23  &75.19$\pm$0.82  \\     
VersBand  & \bf2.65 & \bf 1.19  & \bf90.17$\pm$1.55&\bf83.54$\pm$1.32 \\   
\bottomrule
\end{tabular}
\caption{Results of vocal-to-song generation. GT means GT vocal.}
\label{tab: v2s}                 
\end{table*}

\begin{table}[t]
\small
\centering
\begin{tabular}{l|ccc}
\toprule
\bf Methods            & MOS-Q$\uparrow$   & FEE$\downarrow$   \\
\midrule
VocalBand (w/o GT)           & 4.04$\pm$0.08 & 0.07     \\
\midrule
StyleSinger           & 3.79$\pm$0.10  & 0.09  \\
VocalBand  & \bf3.87$\pm$0.05  & \bf0.08   \\
\bottomrule 
\end{tabular}
\caption{Results of accompaniment-to-song generation. GT means GT accompaniment.}
\label{tab: a2s}
\end{table}

\subsection{Vocal Generation}
\label{app: svs}

In Figure \ref{fig: svs}, we compare the mel-spectrograms of VocalBand with different singing styles specified in the text prompt. 
Figure \ref{fig: svs} (a) represents the GT vocal, where the mel-spectrogram within the yellow box is relatively uniform, indicating a stable vocal performance, while the F0 contour in the red box is smooth. 
In contrast, Figure \ref{fig: svs} (b) does not specify singing styles, allowing the free use of techniques to enhance expressiveness, as seen by the significant pitch oscillations in the red box, characteristic of vibrato. 
In Figure \ref{fig: svs} (c), representing the breathy technique, the mel energy in the yellow box shows a significant drop in high-frequency energy, consistent with the softer, airier vocal timbre of breathy singing.
Finally, Figure \ref{fig: svs} (d) illustrates the bubble technique, where the yellow box displays pronounced low-frequency energy with more exaggerated vertical modulations. 
The red box shows a distinctive pitch fluctuation pattern, characterized by slower, larger oscillations, indicative of the unique vocal fold vibrations in this technique.
These results demonstrate that VocalBand can achieve diverse and highly expressive control over the same content based on the different singing styles specified in the text prompt.

\subsection{Singing Style Transfer}
\label{app: st}

In Figure \ref{fig: st}, we compare the performance of VocalBand and baseline models on singing style transfer. 
It can be observed that VocalBand excels at capturing the intricate nuances of the prompt style. 
The pitch curve generated by VocalBand displays a greater range of variations and finer details, closely resembling the prompt style.
In the yellow boxes, it is evident that VocalBand captures nuances in pronunciation and articulation skills similar to the vocal prompt. 
In contrast, the curves generated by other methods appear relatively flat, lacking distinctions in singing styles.
Moreover, the mel-spectrograms generated by VocalBand exhibit superior quality, showcasing rich details in frequency bins between adjacent harmonics and high-frequency components. 
In contrast, the mel-spectrograms produced by other methods demonstrate lower quality and a lack of intricate details.

\subsection{Music Style Transfer}

For music style transfer, AccompBand uses the noisy prompt accompaniment $\tilde{y_a}$ with a time step 0.5 instead of Gaussian noise $\epsilon$ and sums it with the target vocal $y_v$, enabling the model to learn the style from the retained components of the prompt accompaniment. 
Thus, we do not need a text prompt to control the music style. 
We use ALI-A for subjective evaluation of the style similarity to the prompt accompaniment.
As shown in Table \ref{tab: mst}, we achieve good style similarity with minimal changes in quality. 
This demonstrates that VersBand, leveraging AccompBand's flow matching mechanism, can also effectively perform the music style transfer task.

\subsection{Vocal-to-Song Generation}

We can directly input GT vocals for the vocal-to-song generation task. 
We compare our method with MelodyLM, which also generates songs from GT vocals. 
We use the objective metrics reported in their papers and subjectively evaluate the demos on their demo pages.
As shown in Table \ref{tab: v2s}, it is evident that with GT vocal input, VersBand achieves improved quality and better alignment with the vocals compared to song generation without GT vocal input, and it outperforms MelodyLM. 
This is because the GT vocal provides a more accurate style, melody, and rhythm, better matching the target accompaniment.
It demonstrates that VersBand effectively utilizes AccompBand's excellent vocal alignment mechanisms of Aligned MOE, to accomplish the Vocal-to-Song Generation task.

\subsection{Accompaniment-to-Song Generation}

We use ROSVOT \citep{li2024robust} to extract notes from the accompaniment to guide VocalBand for vocal generation. 
The extracted notes are also provided to StyleSinger, which can similarly utilize notes, as a baseline model.
As shown in Table \ref{tab: a2s}, it is evident that the quality decreases when using GT accompaniment instead of music scores, as the notes from the accompaniment are not aligned with the vocal notes, primarily due to differences in their characteristics. 
Vocals often involve techniques and emotional expression, with pauses between words. 
At the same time, accompaniments are more complex, involving multiple instruments and rarely pausing, leading to discrepancies in timing and pitch between the vocal and accompaniment notes.
However, VocalBand still outperforms StyleSinger and achieves satisfactory results. 
This demonstrates that VersBand can leverage the user's preferred GT accompaniment for vocal pairing, with VocalBand exhibiting excellent rhythm and melody control by decoupling content.

\end{document}